\documentclass[a4paper,structabstract]{aa}

\usepackage{float}
\usepackage{mathenv}
\usepackage{aas_macros}
\usepackage{natbib}
\usepackage{txfonts}
\usepackage{array}
\usepackage{multirow} 
\usepackage[]{caption}
\usepackage{wasysym} 
 \usepackage[squaren,Gray]{SIunits}
\usepackage[rightcaption]{sidecap}
\sidecaptionvpos{figure}{c} 
\usepackage{graphicx}

 \usepackage{mathtools} 
 \usepackage{booktabs} \newcommand{\ra}[1]{\renewcommand{\arraystretch}{#1}}

\usepackage{hyperref}

\hypersetup{
colorlinks=true,
citecolor=blue,
linkcolor=blue,
urlcolor=black
}

\def\wig#1{\mathrel{\hbox{\hbox to 0pt{%
          \lower.6ex\hbox{$\sim$}\hss}\raise.4ex\hbox{$#1$}}}}

\def\ie{{\rm i.e.,}}  
\def\eg{{\rm e.g.,}}  

\def\teff{T_{\rm eff}}

\def\tint{T_{\rm int}}
\def\tirr{T_{\rm irr}}
\def\tmu{T_{\rm{\mu_{*}}}}

\def\fh{f_{\rm H}}
\def\D{\mathrm{d}}

\def\Ka{\kappa_{\rm 1}}
\def\Kb{\kappa_{\rm 2}}

\def\Ga{\gamma_{\rm 1}}

\def\Gp{\gamma_{\rm p}}
\def\Gv{\gamma_{\rm v}}

\def\Gva{\gamma_{\rm v1}}
\def\Gvb{\gamma_{\rm v2}}
\def\Gvc{\gamma_{\rm v3}}
\def\betav{\beta_{\rm v}}

\def\D{\mathrm{d}}
\def\K{\mathrm{K}}
\def\teffo{T_{\rm eff_{0}}}

\def\tt{}

\defcitealias{Parmentier2014a}{Paper~I}

\begin{document}

\title{A non-grey analytical model for irradiated atmospheres.\thanks{A FORTRAN implementation of the analytical model is available in electronic form at the CDS via anonymous ftp to cdsarc.u-strasbg.fr (130.79.128.5) or via http://cdsweb.u-strasbg.fr/cgi-bin/qcat?J/A+A/ or at http://www.oca.eu/parmentier/nongrey.}}
\subtitle{II: Analytical vs. numerical solutions}
\titlerunning{A non-grey analytical model for irradiated atmospheres II}
\authorrunning{Parmentier}

\author{Vivien Parmentier\inst{1,2,3}
\and
Tristan Guillot\inst{1,2}
\and
Jonathan J. Fortney\inst{2}
\and
Mark S. Marley\inst{4}}

\offprints{V.Parmentier}

\institute{
Laboratoire J.-L. Lagrange, Universit\'e de Nice-Sophia Antipolis, CNRS, Observatoire de la C\^ote d'Azur, BP 4229, 06304 Nice,
France \email{vivien.parmentier@oca.eu}\\
\and 
Department of Astronomy and Astrophysics, University of California, Santa Cruz, CA 95064, USA
\and
NASA Sagan Fellow
\and NASA Ames Research Center, MS-245-3, Mofett Field, CA 94035
}

\date{Accepted for publication in A\&A 29/10/2014}

\abstract
{The recent discovery and characterization of the diversity of the atmospheres of exoplanets and brown dwarfs calls for the development of fast and accurate analytical models.}
{We wish to assess the goodness of the different approximations used to solve the radiative transfer problem in irradiated atmospheres analytically, and we aim to provide a useful tool for a fast computation of analytical temperature profiles that remains correct over a wide range of atmospheric characteristics. }
{We quantify the accuracy of the analytical solution derived in paper I for an irradiated, non-grey atmosphere by comparing it to a state-of-the-art radiative transfer model. Then, using a grid of numerical models, we calibrate the different coefficients of our analytical model for irradiated solar-composition atmospheres of giant exoplanets and brown dwarfs.}
{We show that the so-called Eddington approximation used to solve the angular dependency of the radiation field leads to relative errors of up to $\sim 5\%$ on the temperature profile. For grey or semi-grey atmospheres (\ie when the visible and thermal opacities, respectively, can be considered independent of wavelength), we show that the presence of a convective zone has a limited effect on the radiative atmosphere above it and leads to modifications of the radiative temperature profile of approximately $\sim 2\%$. However, for realistic non-grey planetary atmospheres, the presence of a convective zone that extends to optical depths smaller than unity can lead to changes in the radiative temperature profile on the order of $20\%$ or more. When the convective zone is located at deeper levels (such as for strongly irradiated hot Jupiters), its effect on the radiative atmosphere is again of the same order ($\sim 2\%$) as in the semi-grey case. {\tt We show that the temperature inversion induced by a strong absorber in the optical, such as TiO or VO is mainly due to non-grey thermal effects reducing the ability of the upper atmosphere to cool down rather than an enhanced absorption of the stellar light as previously thought.} \textnormal{Finally, we provide a functional form for the coefficients of our analytical model for solar-composition giant exoplanets and brown dwarfs. This leads to fully analytical pressure--temperature profiles for irradiated atmospheres with a relative accuracy better than $10\%$ for gravities between $2.5\,\rm m\,s^{-2}$ and $250\,\rm m\,s^{-2}$ and effective temperatures between $100\,K$ and $3000$\,K. This is a great improvement over the commonly used Eddington boundary condition.}}
{}


\maketitle
%

\section{Introduction}

The large diversity of exoplanets in terms of irradiation temperature, gravity, and chemical composition discovered around stars with different properties calls for the development of fast, accurate, and versatile atmospheric models.

In paper~I \citep{Parmentier2014a}, we derived a new analytical model for irradiated atmospheres. Unlike previous models, our model takes into account non-grey opacities both in the visible and in the thermal frequency ranges.  
Using two different opacity bands in the thermal frequency range, we highlighted the dual role of thermal non-grey opacities in shaping the thermal structure of the atmosphere. Opacities dominated by lines (\ie~opacities where the lowest of the two values is dominant) enable the upper atmosphere to cool down significantly compared to a grey atmosphere whereas opacities dominated by bands (\ie~opacities where the highest of the two values is dominant) lead to a significant cooling of the upper atmosphere and to a significant heating of the deep atmosphere.

The pressure and temperature dependent line-by-line opacities that are used in numerical models to compute accurate temperature profiles are represented in analytical models by only a handful of parameters. Thus, to compute accurate temperature structure from our analytical model for specific planetary atmospheres, we need to know how those parameters vary with the physical properties of the planet. 

In this study, we apply our model to irradiated, solar-composition, semi-infinite atmospheres \eg~brown dwarfs, giant planets or planets with a surface situated in the optically thick region of the atmosphere. Based on the results from a state-of-the-art numerical model, we assess the goodness of the different approximations inherent in analytical solutions of the radiative transfer equation. Then, using a grid of numerical models, we calibrate the different coefficients of our analytical model and provide a useful tool for a fast computation of analytical temperature profiles for planet atmospheres that remains correct over a wide range of gravity and irradiation temperatures.

As a first step, in Sect.~\ref{sec::Eddington} we quantify the accuracy of models derived with the Eddington approximation, a common simplification of the radiative transfer equation in analytical model atmospheres. Then in Sect.~\ref{sec::Convection} we build a simple radiative/convective model where the radiative solution of~\citetalias{Parmentier2014a} is replaced by a convective solution whenever the Schwarzschild criterion is verified. We then discuss and quantify the intrinsic error of such a simple model of convective adjustment. Then, guided by a state-of-the art numerical integration of the radiative transfer equation, we constrain the parameters of the analytical solution of~\citetalias{Parmentier2014a} to develop a fully analytical solution for the atmospheric temperature/pressure profiles of irradiated giant planets. The solution presented in Sect.~\ref{sec::Results} reproduces with a $10\%$ accuracy the numerical solutions over a wide range of gravity and irradiations. {\tt Finally, in Sect.~\ref{sec::TiO} we highlight the important role of non-grey thermal effects in the influence of titanium oxide on the temperature profiles of irradiated planets.}
\section{Models}
\label{sec::models}
{\tt
\subsection{Setting}
\label{sec::Setting}
We consider the case of a planet with a thick atmosphere (\ie~a planet with no surface or with a surface at very high optical depth) orbiting at a distance $a$ from its host star of radius $R_{*}$ and effective temperature $T_{*}$. The total flux received by the planet is $4\pi R_{\rm p}^{2}\sigma T_{\rm eq_{0}}$ where the equilibrium temperature for zero albedo is defined as
\begin{equation}
T_{\rm eq_{0}}^{4}\equiv \frac{T_{*}^{4}}{4}\left(\frac{R_{*}}{a}\right)^{2}.
\end{equation}
At a given point in the planet, the incoming stellar flux is characterized by a temperature
\begin{equation}
T_{\rm irr_{0}}^{4}=4T_{\rm eq_{0}}^{4}.
\end{equation}
This flux hits the atmosphere with an angle $\theta_{*}$ with the vertical direction. We define $T_{\rm \mu_{*0}}$ as the projected flux hitting the top of the atmosphere:
\begin{equation}
T_{\rm \mu_{*0}}^{4}=\mu_{*}T_{\rm irr_{0}}^{4}\,.
\end{equation}
Where $\mu_{*}=\cos{\theta_{*}}$. A part $1-A_{\mu_{*}}$ of this flux is reflected back to space, where $A_{\mu_{*}}$ is the angle-dependent reflectivity. We characterize the flux that penetrates the atmosphere by the temperature $T_{\mu_{*}}$
\begin{equation}
T_{\mu_{*}}^{4}=\mu_{*}T_{\rm irr}^{4}\,.
\end{equation}
where $T_{\rm irr}^{4}=(1-A_{\mu_{*}})T_{\rm irr_{0}}^{4}$. 

While these definitions are useful to calculate the temperature profile at a given location in the atmosphere, they need to be averaged over $\mu_{*}$ to calculate the mean state of the atmosphere. As shown by~\citet{Guillot2010}, the temperature profile obtained for a mean stellar angle of $\mu_{*}=1/\sqrt{3}$ and an incoming flux equal to a fraction of the total incoming flux is a reasonable approximation of the exact mean temperature profile of the planet. When considering average profiles we use the so-called \emph{isotropic approximation}~\citep[\eg][]{Guillot2010}
\begin{equation}
 \left\{
\begin{array}{l l}
 \mu_{*}=1/\sqrt{3}\\
  T_{\mu_{*}}^{4}=(1-A_{\rm B})4fT_{\rm eq_{0}}^{4}\\ 
\end{array}\, \right. ,
\end{equation}
where $A_{\rm B}$ is the Bond albedo of the planet. The relationship between $A_{\rm B}$ and $A_{\mu_{*}}$ is not straightforward and will be discussed in more details in Sect.~\ref{sec::Albedos}. $f$ is a parameter smaller than one. For $f=0.25$ we have $T_{\mu_{*}}=T_{\rm eq}$ and the thermal profile is close to the planet average profile. For $f=0.5$, the thermal profile corresponds to the dayside average profile. The average profiles are obtained by setting $\mu_{*}T_{\rm irr}^{4}=T_{\mu_{*}}^{4}$ and $\mu_{*}=1/\sqrt{3}$ in equation (76) of~\citetalias{Parmentier2014a}\footnote{A factor $\mu_{*}^{1/4}$ was missing in the expression of $T_{\rm irr}$ in section 3.7 of ~\citetalias{Parmentier2014a}.}.

The atmosphere is also heated by the planet interior and receives a flux $\sigma T_{\rm int}^{4}$ from below. The total energy budget of the atmosphere is characterized by its effective temperature
\begin{equation}
T_{\rm eff}^{4}=T_{\mu_{*}}^{4}+T_{\rm int}^{4}\,,
\end{equation}
which is valid for both the averaged and the non-averaged cases\footnote{The last equation of the first footnote of~\citetalias{Parmentier2014a} should thus read $\teff^{4}=\tint^{4}+\tirr^{4}/\sqrt{3}$.}. Finally, we define an effective temperature for zero albedo :
\begin{equation}
T_{\rm eff_{0}}=T_{\mu_{*0}}^{4}+T_{\rm int}^{4}\,.
\end{equation}
All the quantities defined for zero albedo, denoted by the subscript $0$ are independent of the properties of the atmosphere and can be calculated \emph{a-priori}. All other quantities are characteristic of a given atmosphere and need to be constrained either by the observations or determined by numerical calculations.
 }
\subsection{Opacities}
\label{sec::Opacities}
The interaction between photons and atmospheric gas is described by opacities which are a function of the wavelength of the radiation and of the temperature, pressure and composition of the gas. Although the variety of mixtures and cases to be considered is infinite, we choose to limit the present study to one set of opacities because of its very extensive use both in the context of giant exoplanets and brown dwarfs, \ie the solar-composition opacities provided by \cite{Freedman2008}. These opacities have been calculated for a solar-composition mixture in chemical equilibrium. They do not account for the presence of clouds, and any chemical species that condenses at a given temperature and pressure is taken out of the mixture. Although clouds are thought to exist in planetary atmospheres \citep[see][for a review]{Marley2013} and should affect the thermal structure of their atmosphere~\citep[\eg][]{Heng2012}, we do not take into account scattering by cloud particles in this study. The first order effect of clouds is to reflect part of the incoming stellar light to the space, what can be taken into account by specifying the relevant Bond Albedo when calculating the effective temperature of the model. {\tt Scattering by the gas, however, is taken into account.}

\begin{figure}
\includegraphics[width=\linewidth]{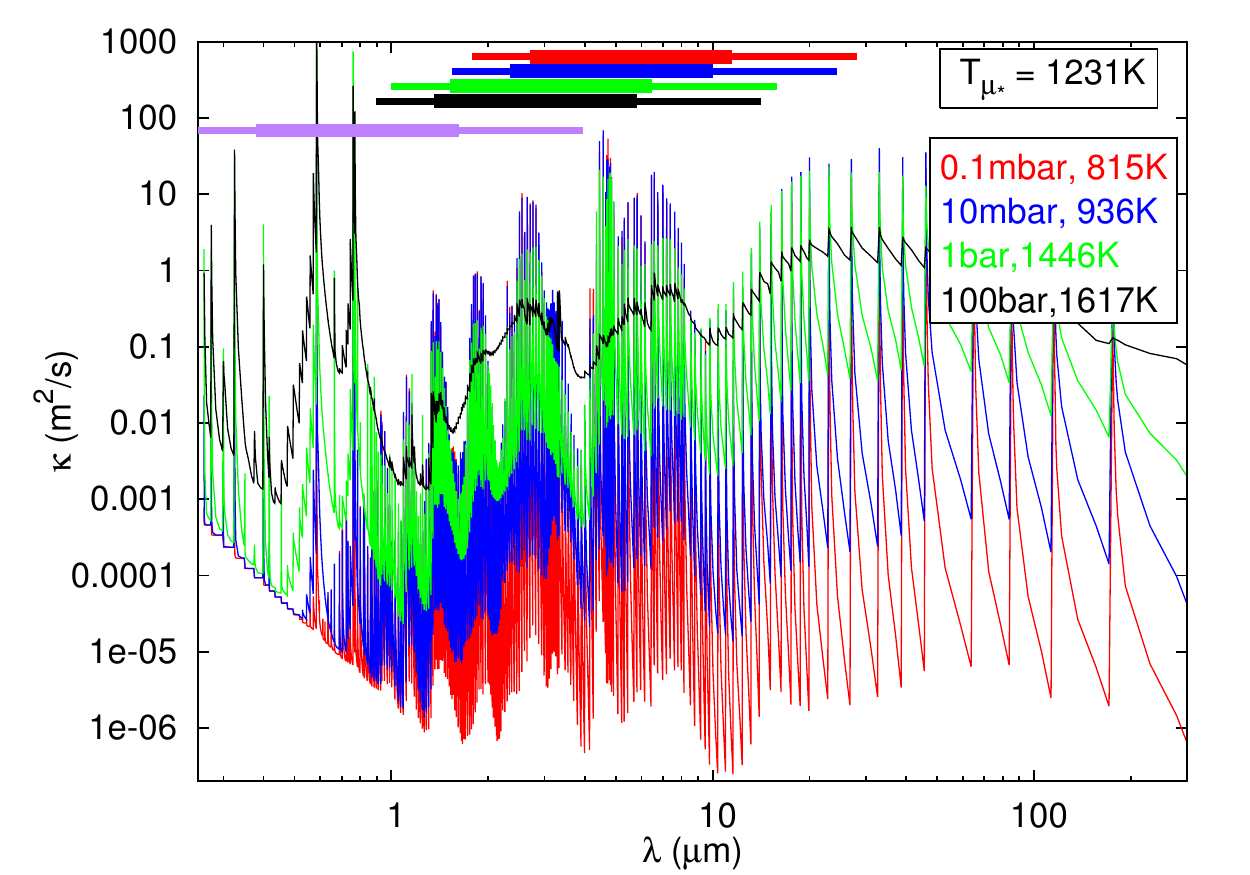}
\caption{line-by-line opacities as a function of wavelength for five different conditions corresponding to different points in the PT profile of a giant planet with $g=\unit{25}\meter\per\second\squared$, $\mu_{*}=1/\sqrt{3}$, $\tint=\unit{100}K$ and $\tmu=\unit{1231}K$, corresponding to the dayside average profile of a planet orbiting at $0.053\,{\rm AU}$ from a sun-like star. Inside each bin of frequency, we plot the cumulative distribution function of the opacities instead of the line-by-line opacity function. The purple bar represents the wavelength range where $90\%$ (thick line) and $99\%$ (thin line) of the stellar energy is emitted. The other horizontal bars represents the wavelength range of the thermal emission of the planet at different locations along the PT profile. }
\label{fig::Opacities}
\end{figure}

While tens of millions of lines have been used for the calculation of these opacities, we choose to show them in Fig.~\ref{fig::Opacities} in the same form as they are used by the numerical code described hereafter in Section~\ref{sec:egp}: In the so-called correlated-$k$ method, the opacities values are sorted from the lowest to the highest values within a limited number of spectral bins (in our case 196), {\tt this is similar to the opacity distribution function (ODF) method in the stellar atmospheres modeling~\citep{Strom1966}}. As long as the spectral bins are small compared to the width of the local Planck function, the error made on the wavelength corresponding to a given opacity is expected to be small and the consequences for the computed temperature profile limited \citep[see][]{Goody1989}. Figure~\ref{fig::Opacities} shows the opacities for different pressure and temperature points taken along a selected planetary temperature/pressure profile corresponding approximately to a solar composition 1-Jupiter mass and radius planet at 0.05\,AU from a Sun-like star. The wavelength range in which the Planck function has 90\% and 99\% of the total energy is shown by the thick and thin horizontal bars, respectively, for the different temperatures considered. The contribution of the spectral lines to the opacities shapes the cumulative distribution function inside each bin. As one moves progressively from the top to the bottom of the atmosphere, pressure (always) and temperature (generally) increase which broadens the spectral line profiles. This results in a flattening of the cumulative opacity distribution function within each bin. Opacities in the upper atmosphere are characterized by very strong variations with wavelength and a comb-like structure. Deeper-down, the wavelength dependence is mostly due to the presence of molecular bands and takes place on scales significantly larger than our bin size.

An important feature of these opacities is that most of the variations of the opacity with wavelength take place on scales shorter than the characteristic wavelength range of the Planck function. This is certainly the case at low-pressures when the opacity varies extremely quickly with wavelength, but it remains true (to some extent) at high pressures in the band regime. Another feature of irradiated atmospheres is that the temperature variations remain limited so that there is always a significant overlap between the Planck function from the low to the high optical depth levels. These two features justify the use of the picket-fence approximation, and hence of the analytical model of~\citetalias{Parmentier2014a}.

\subsection{Analytical model}
\label{sec::Analytical}

Although analytical models of irradiated atmospheres can only be obtained for very restrictive approximations on the opacities, they provide nonetheless a useful tool to understand the physics of the radiative transfer and to compute with a low computational cost temperature profiles for a large variety of atmospheric properties. In the particular model derived in~\citetalias{Parmentier2014a} the line-by-line opacities are modeled by two different homogeneous set of lines, the full opacity function being described by six independent parameters.

The first set of lines, described by three parameters, represents the thermal part of the opacities, \ie~the part of the opacity function in the frequency range covered by the local Planck function of the atmospheric thermal emission. The Rosseland mean opacity $\kappa_{\rm R}(P,T)$ is the only one of those parameters that can vary with depth in the atmosphere. In particular, it is the relevant opacity to describe accurately the energy transport in the optically thick part of the atmosphere~\citep{Mihalas1978,Mihalas1984}. The other two parameters describe the non-greyness of the opacities, $\ie$ their variation in frequency. The first one, $\Gp$ is the ratio of the Planck mean opacity to the Rosseland mean opacity, where the Planck mean opacity is dominated by the highest values of the opacities whereas the Rosseland mean is dominated by he lowest values of the opacities. Thus, grey opacities have $\Gp=1$ and any departure from the grey model increases $\Gp$. The second parameter, $\beta$, is the relative width of the opacity lines. Values of $\beta$ lower than $0.1$ represents opacities dominated by atomic lines whereas values of $\beta$ between $0.1$ and $0.9$ correspond to opacities dominated by molecular bands. In the following sections, the parameter $\Gp$ will sometimes be replaced by an equivalent parameter : $\kappa_{\rm 1}/\kappa_{\rm 2}$, where $\kappa_{\rm 1}$ is the highest of the two opacities and $\kappa_{\rm 2}$ the lowest. Value of $\kappa_{\rm 1}/\kappa_{\rm 2}$ between $10^{4}-10^{5}$ in the upper atmosphere and between $10-100$ in the deep atmosphere can be estimated from Fig.~\ref{fig::Opacities} for a typical hot-Jupiter. The simple relationship between $\kappa_{\rm 1}/\kappa_{\rm 2}$ and $\Gp$ is described by Eq.~(87) of~\citetalias{Parmentier2014a}.

{\tt The second set of lines represents the visible parts of the opacities, \ie~the part of the opacity function in the frequency range covered by the Planck function of the stellar irradiation. Since the planet's atmosphere is usually cooler than the stellar photosphere, the two set of opacity lines can be considered independent of each other. Whereas the model cannot take into account more than two thermal opacity bands, it can model as many visible bands as needed. Here we choose to represent the visible opacities with three different bands of adjustable strength represented by $\Gva$, $\Gvb$ and $\Gvc$ the ratio of the highest, medium and lowest opacity to the thermal Rosseland mean opacity. Each band is supposed to have the same spectral width described by $\beta_{\rm v1}=\beta_{\rm v2}=\beta_{\rm v3}=1/3$.}
{\tt
Although we differentiate the \emph{visible} and the \emph{thermal} opacities throughout the paper, the difference is not based on a spectral difference but rather on a geometrical one\footnote{Although the visible opacities are usually in the optical spectral range and the thermal opacities in the infrared spectral range, it is not necessarily the case.}. As shown in~\citetalias{Parmentier2014a} the stellar flux is a collimated beam propagating downward in the atmosphere and can be traced down until it is absorbed by the atmosphere. At each level, it's absorption is proportional to the remaining flux times the local opacities. The mean opacities relevant to understand the absorption of the stellar flux at a given atmospheric level are therefore a combination of the spectrally dependent remaining flux and opacities at this level. The other part of the radiation, sometimes called the \emph{diffuse} component has a more complex geometry that is often approximated via the Eddington approximation (see Sect.~\ref{sec::Eddington}). The thermal opacities characterize how the diffuse radiation is absorbed and emitted. Both visible and thermal radiation can be related as they both depend on the physical and chemical properties of the atmosphere but do not have to be equal, even when the stellar flux and the planetary emission overlap in wavelength. 
}

In our analytical model, the Rosseland mean opacity can vary with pressure and temperature. Thus, physical processes producing an overall increase in the opacities, such as the increasing importance of the collision induced absorption with pressure can be accurately taken into account. Our model is the first analytical model to take into account non-grey thermal opacities in irradiated atmosphere. However, the variation of the opacity with frequency cannot change through the atmosphere. Therefore, all the other coefficients must remain constant in the whole atmosphere and a physical phenomenon such as the variation of the pressure or thermal broadening of the lines through the atmosphere cannot be taken into account.

\subsection{Numerical model}\label{sec:egp}

Whereas analytical models are confined to model atmospheres with very simplified opacities, the radiative transfer equation can be solved by numerical integration using the full, line-by-line, frequency, pressure-and temperature-dependent opacities described in Sect.~\ref{sec::Opacities}. Moreover, numerical models can integrate the radiative transfer equation by taking into account an arbitrary high number of angular directions, with no need to invoke the Eddington approximation. 

Here, we use the EGP (Extrasolar Giant Planet) code initially developed by~\citet{McKay1989} for the study of Titan's atmosphere. Since then, it has been extensively modified and adapted for the study of giant planets \citep{Marley1999}, brown dwarfs \citep{Marley1996,Marley2002,Burrows1997}, and hot Jupiters \citep[\eg][]{Fortney2005a, Fortney2008, Showman2009}.
The version of the code we employ solves the radiative transfer equation using the the delta-discrete ordinates method of \citet{Toon1989} for the incident stellar radiation and the two-stream source function method, also of \citet{Toon1989}, for the thermal radiative transfer. In some cases incident stellar and emitted thermal radiation bands may overlap, but the radiative transfer is solved separately for each radiation source. Opacities are treated using  the correlated-k method \citep[\eg][]{Goody1989}. We consider 196 frequency bins ranging from $0.26$ to $\unit{300}\micro\meter$; within each bin, the information of typically 10,000 to 100,000 frequency points is compressed inside a single cumulative distribution function that is then interpolated using 8 $k$-coefficients. The angular dependency is computed using the Gauss quadrature formula for the fluxes. This formula allows us to transform an integral over $\mu=\cos\theta$ into a simple sum over angles
\begin{equation}
\int_{-1}^{1}\mu I_{\nu}(\mu)\,\D\mu=\sum_{i=1}^{n}\omega_{i}I_{\nu}(\mu_{i})
\end{equation}
with the $\omega_{i}$ and the $\mu_{i}$ being tabulated in~\citet{Abramowitz1965}. Here we use 5 Gauss points. The EGP model calculates a self-consistent radiative/convective solution, deriving the adiabatic gradient using the equation of state of~\citet{Saumon1995} but can also look for a fully radiative solution.

Although numerical models were built in order to incorporate the full complexity of the opacity function, it can nonetheless solve the radiative transfer equation with the same simplifications than the ones used in the analytical models. In particular, the $k$-coefficient method can be used to easily implement the simplified opacities of~\citet{Parmentier2013} by setting a given number of $k$-coefficients at $\Ka$ and the other ones at $\Kb$ in each frequency bin. Moreover, the opacities used to compute the absorption of the stellar flux can be independent from the opacities used to compute the thermal fluxes. Simplified opacities as in the analytical case can therefore be used.

\subsection{Comparison to an asymptotically exact solution}
\label{sec::Convergence}
In order to test the validity of the radiative solution found by the numerical model, we compare it to the analytical solution obtained by the method of discrete ordinates in the grey case~\citep{Chandrasekhar1960}. This method solves the moment equations with grey opacities for a non-irradiated atmosphere by replacing the integrals over angle by a Gaussian sum. By increasing the number of terms in the sum (\ie\ the order of the calculation), it converges towards the exact solution. The first order solution being equivalent to the Eddington approximation. 

\begin{figure}[h]
\includegraphics[width=\linewidth]{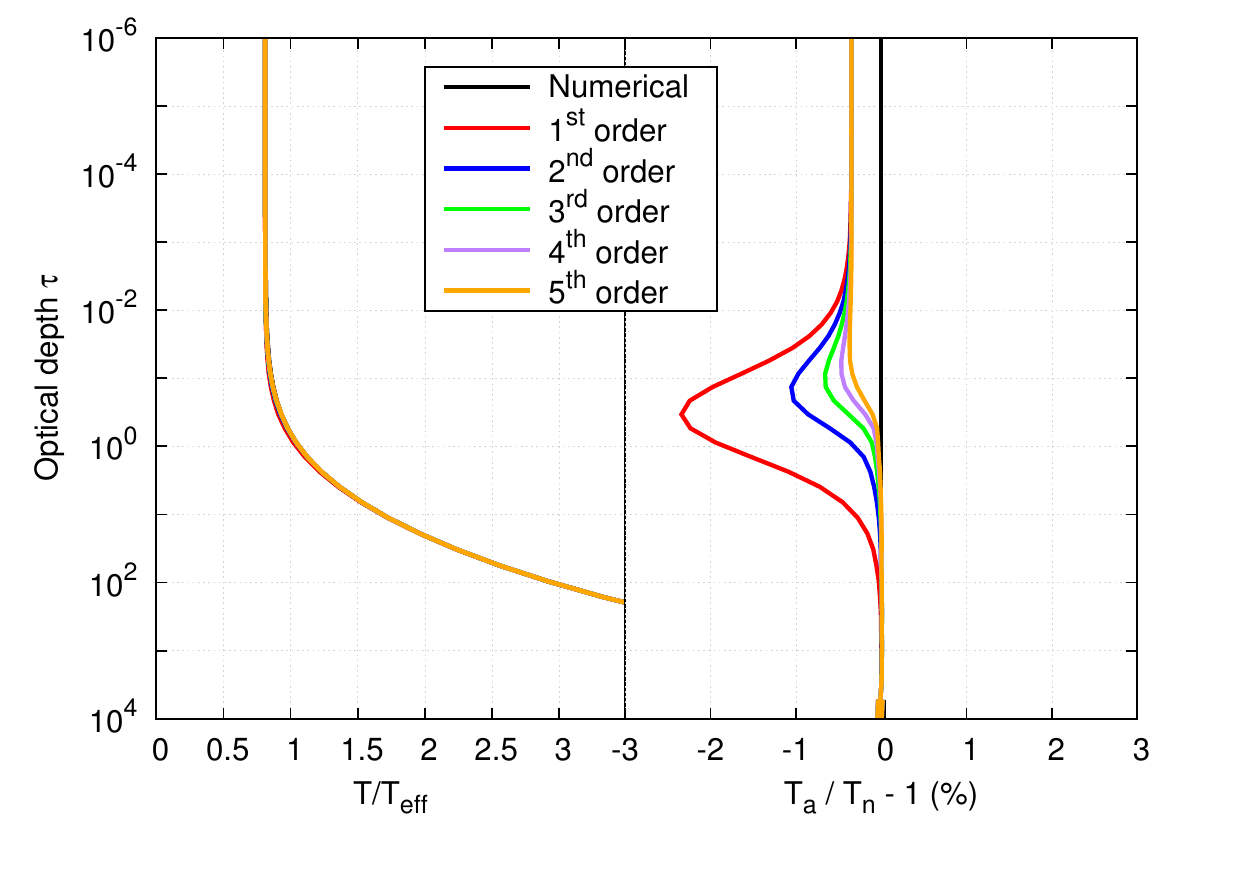}
\caption{Radiative numerical temperature profile in units of effective temperature as a function of optical depth compared to the analytical solution from~\citet{Chandrasekhar1960} using the discrete ordinate in the first, second, third, fourth and fifth approximation. The left panel shows the profiles whereas the right panel shows their relative difference ($T_{\rm a}/T_{\rm n}-1$ where $T_{\rm a}$ is the analytical solution and $T_{\rm n}$ is the numerical solution).}
\label{fig::Convergence-Grey}
\end{figure}

In Fig.~\ref{fig::Convergence-Grey}, we compare the numerical model for the grey, non-irradiated case to these analytical solutions up to the 5th order. The first order analytical solution deviates from the others and from the numerical result by about $2\%$ with the maximum deviation occurring near optical depth unity. We can therefore expect the analytical models based on the Eddington approximation to differ from the exact results by about this value at least -- we will come back to that in Sect.~\ref{sec::Eddington}. The higher order analytical solutions appear to smoothly converge towards the exact solution, but the numerical solution is found to be about $\sim 0.5\%$ warmer at low optical depths. This discrepancy arises from a different use of the Gaussian quadrature formula in the two approaches. Whereas the analytical solution uses the Gaussian quadrature to compute the integral $\int_{-1}^{1}I_{\nu}(\mu)\,\D\mu$, the numerical code uses the quadrature formula to compute the flux integral $\int_{-1}^{1}\mu I_{\nu}(\mu)\,\D\mu$. Therefore, the $5^{th}$ order analytical solution is formally not the same as the five Gauss points numerical model and does not converge toward the same solution. We tested that using eight Gauss points in the numerical model leads to a solution that is correct to $0.1\%$ when compared to the $8^{th}$ order analytical solution. 

Because a $0.5\%$ error is significantly smaller than the other sources of uncertainties in the model (the first one being due to the use of the Eddington approximation) we chose to use the five Gauss points numerical model. We note that this kind of test is unfortunately not possible in the irradiated case (even in the grey approximation) for which no exact analytical solution is known.

\section{Consequences of the Eddington approximation}
\label{sec::Eddington}

We have seen in Sect.~\ref{sec::Convergence} that an asymptotically exact solution of the radiative transfer problem can be found in the grey, non-irradiated case. Unfortunately, such a solution does not exist when accounting for external irradiation. The angle dependency of the radiative transfer problem therefore has to be approximated. Analytical models such as the one in \citetalias{Parmentier2014a} use a closure relation between two moments of the intensity field $I_\nu(\mu)$ (with $\nu$ the frequency of the radiation):
\begin{equation}
\int_{-1}^{1} I_\nu(\mu)\mu^2 \,\D\mu \approx {\frac{1}{3}} \int_{-1}^{1} I_\nu(\mu) \,\D\mu.
\label{eq::Closure}
\end{equation}
This approximation is exact in two specific cases: when the radiation field is isotropic ($I_\nu(\mu)=\rm cte\ \forall \mu$) and when radiation field is semi-isotropic ($I_{\nu}(\mu)=I^{+} \forall \mu>0$ and $I_{\nu}(\mu)=I^{-} \forall \mu<0$). In the deep atmosphere, the radiation is quasi-isotropic and this approximation holds. Toward the top of the atmosphere, most of the thermal radiation comes from the deep layers and is therefore close to be semi-isotropic. In between, the solution is only an approximation. In addition, a boundary condition relating two other moments of the intensity field must be adopted :
\begin{equation}
\int_{-1}^{1} I_\nu(\mu)\mu \,\D\mu\,\bigg|_{top} \approx f_{\rm H}\int_{-1}^{1} I_\nu(\mu) \,\D\mu\,\bigg|_{top}.
\label{eq::boundary}
\end{equation}
These two conditions form what is called the Eddington approximation. 

In the grey, non-irradiated case, those two approximations are linked and $\fh=1/2$. However using equation \ref{eq::Closure} and imposing $\fh=1/\sqrt{3}$ leads to the exact solution at the top of the atmosphere, even though it lacks of self-consistency. In the irradiated case and in the non-grey case the two approximations are independent and $\fh$ is usually set to either $1/2$ or $1/\sqrt{3}$, following the grey, non-irradiated case \citepalias[see][for a complete discussion]{Parmentier2014a}.

As discussed in Sect.~\ref{sec::Convergence}, the relative uncertainty on the temperature profile resulting from the Eddington approximation is $\sim 2\%$ in the grey, non-irradiated case. In order to estimate its magnitude in the grey and non-grey irradiated cases, we must rely on comparison with numerical models. We hereafter adopt the EGP numerical model with 5 Gauss points.

Now we compare the radiative solutions from our numerical model and different analytical models using the simplified opacities described in Sect.~\ref{sec::Analytical}. Thus the solution can be expressed as a function of the Rosseland optical depth $\tau$ only and is independent of the Rosseland mean opacity or of the gravity. Once normalized by the effective temperature, the temperature as a function of the optical depth in each model only depends on the values of $\Gv$,$\Gp$ (or $\kappa_2/\kappa_1$), $\beta$ and the ratio $\tirr/\tint$.

\subsection{Irradiated semi-grey solutions}\label{sec::semi-grey}

In the semi-grey case, several analytical models have been developed~\citep[\eg][]{Hansen2008,Guillot2010,Robinson2012}. As reviewed in~\citetalias{Parmentier2014a}, those models differ mainly by their choice of $\fh$ and their choice of the upper boundary condition. For simplicity, we will compare only three of them: the two different versions of equation (27) of~\citet{Guillot2010} (with $f_{H}=1/2$ or $f_{H}=1/\sqrt{3}$) and the semi-grey limit of the model derived in~\citetalias{Parmentier2014a} with $\fh=1/2$ which uses as upper boundary condition a mix between the model of~\citet{Guillot2010} and the one of~\citet{Hansen2008}. We compare those models for two different values of the main parameter of semi-grey models: the ratio of the visible to the thermal opacities, $\Gv$.

\begin{figure}[htbp]
\includegraphics[width=\linewidth]{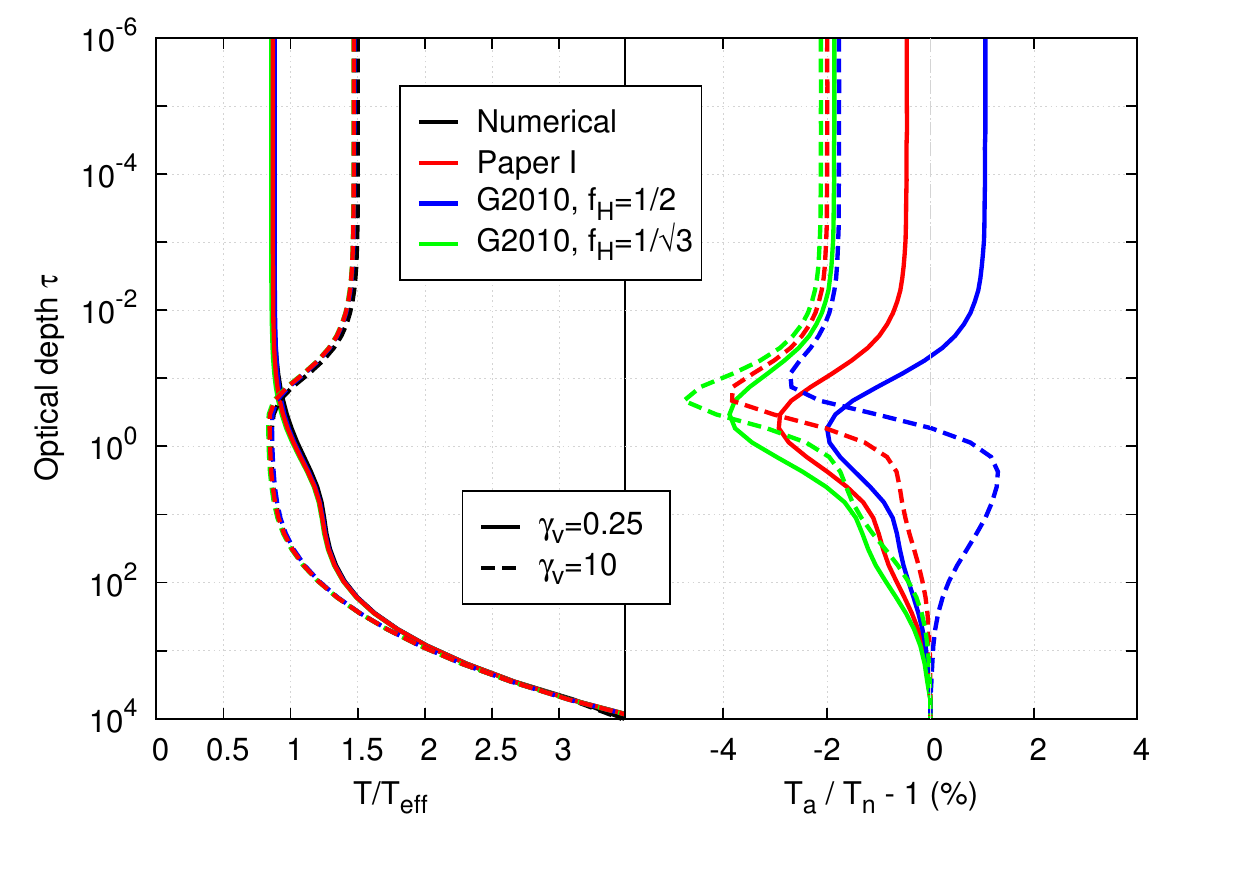}
\caption{Comparison between the radiative numerical solution (black line), our work (red line) and \citet{Guillot2010} model for two different values of $f_{\rm H}$ (blue and green lines) for a fully radiative semi-grey atmosphere with $\Gv=0.25$ (plain lines) or $\Gv=10$ (dashed lines). We set $\mu_{*}=1/\sqrt{3}$, $T_{\rm irr}=\unit{1288}K$ and $T_{\rm int}=\unit{500}K$.
}
\label{fig::Eddington-Irr}
\end{figure}

Figure~\ref{fig::Eddington-Irr} compares these models for a typical irradiated Jupiter-mass exoplanet close to a solar-type star and shows the magnitude of the error which is due to the Eddington approximation -- both the closure relation defined by Eq.~\eqref{eq::Closure} and the adopted value of $\fh$ -- and the chosen upper boundary condition, different between \citet{Guillot2010} and \citetalias{Parmentier2014a}. The left panel shows the temperature profiles as a function of optical depth which mainly depends on the magnitude of the greenhouse effect: when $\Gv$ is small, most of the incoming irradiation is absorbed deep in the atmosphere, the temperature increases monotonously with increasing depth, and the solution behaves like the non-irradiated solution with the same effective temperature (see the $1^{st}$ order case of Fig.~\ref{fig::Convergence-Grey}). When $\Gv$ is large, most of the incoming stellar light is absorbed high up, creating a temperature inversion around visible optical depth unity (and thus thermal optical depth $\tau=1/\Gv$). 

The right panel of Fig.~\ref{fig::Eddington-Irr} shows that the magnitude of the difference between the numerical solution and the analytical ones strongly depends on the choice of $\fh$ and of the top boundary condition, but remains of the same order-of-magnitude as for the non-irradiated grey case of Sect.~\ref{sec::Convergence}. Specifically, the Eddington approximation is found to lead to a $\sim 4\%$ uncertainty on the temperature profile and always converges towards zero at large optical depths. Except for the $f_H=1/\sqrt{3}$ solution, all other analytical solutions, including the one from \citetalias{Parmentier2014a}, systematically underestimate the temperature at a given depth. 

It is obvious from Fig.~\ref{fig::Eddington-Irr} that, unlike the non-irradiated case, no choice of $\fh$ can yield an exact skin temperature $T(\tau=0)$ \citepalias[see related discussion in][]{Parmentier2014a}. 
\subsection{Irradiated non-grey solutions}
We now test the analytical model of~\citetalias{Parmentier2014a} in the non-grey case. In order to do so, we compare the analytical model to the numerical model for different values of the ratio of the thermal opacities $\kappa_2/\kappa_1$ and a single visible channel to the numerical model with the same thermal and visible opacities. We adopt $\beta=0.86$ and $\gamma_{\rm v}=0.25$, typical values needed to reproduce detailed models of hot Jupiters (see Sect.~\ref{sec::Results} hereafter) and the same irradiation and internal temperature as in the previous section. 

\begin{figure}[htbp]
\includegraphics[width=\linewidth]{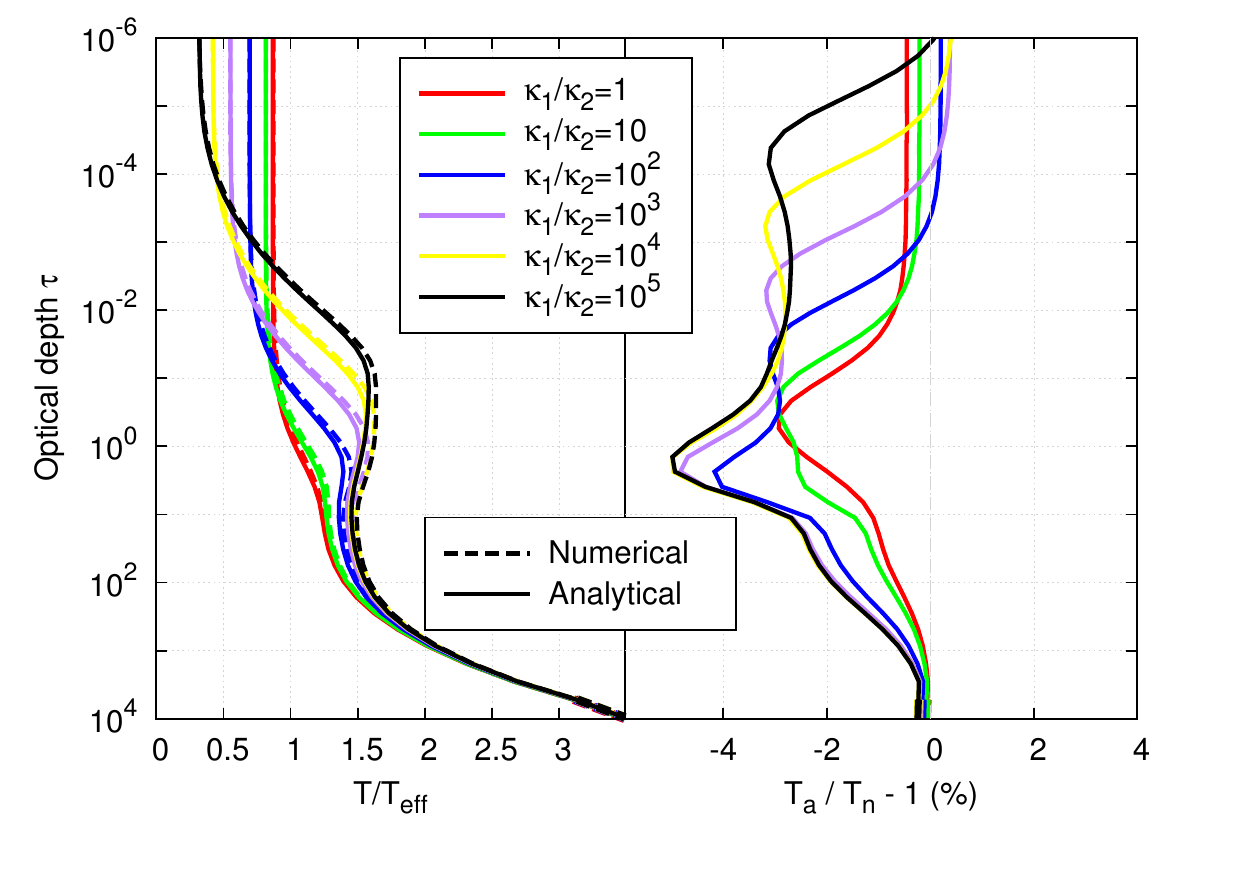}
\caption{Comparison between the analytical model (plain lines) and the radiative numerical model (dashed lines) for different values of $\Kb/\Ka$ (left panel). The right panel shows the relative difference between the analytical and the numerical solution for each case. We used $\mu_{*}=1/\sqrt{3}$, $T_{\rm irr}=\unit{1288}K$, $T_{\rm int}=\unit{500}K$, $\Gv=0.25$ and $\beta=0.86$.}
\label{fig::Eddington-IrrNG}
\end{figure}

Figure~\ref{fig::Eddington-IrrNG} shows the resulting temperature-optical depth profiles and the relative difference between the numerical and analytical solutions. As $\kappa_2/\kappa_1$ increases, the temperature profile gets cooler in the upper atmosphere and warmer in the deep atmosphere, an effect described in details in~\citetalias{Parmentier2014a}. 

The red curve ($\kappa_2/\kappa_1=1$) corresponds to the semi-grey solution already seen in Sect.~\ref{sec::semi-grey} and Fig.~\ref{fig::Eddington-Irr}. As shown in the right panel, the discrepancy between the analytical and numerical models increases with the non-greyness of the opacities. The maximum error (in absolute terms) increases from about $\sim 2\%$ to a little bit less than $\sim 5\%$ when $\kappa_2/\kappa_1$ is increased from $1$ to $10^5$. Moreover, the optical depth range for which the discrepancy is larger than $1\%$ increases at the same time from $[0.1\sim 10]$ to $[10^{-5}\sim 10^{3}]$. 

This increase in the extent of the region in which the temperature profile departs from the numerical solution is a direct consequence of the Eddington approximation in the two thermal channels with opacities $\kappa_1$ and $\kappa_2$, respectively. At high optical depth, in the diffusion limit, the radiation field is isotropic in each thermal channel and the Eddington approximation is valid. At very low optical depth radiation comes mostly from the levels where the first and the second thermal channels become optically thin, much deeper in the atmosphere. Therefore radiation in the optically thin layers is close to be semi-isotropic which validates the choice of the Eddington approximation. Inbetween, the difference between the analytical and the numerical solutions exhibits two maxima. Those maxima correspond to the levels where the first and the second thermal bands become optically thin. As the ratio $\Ka/\Kb$ increases, the first channel becomes optically thin at higher Rosseland optical depth and the second channel becomes optically thin at lower Rosseland optical depth, creating the two-peak feature of Fig.~\ref{fig::Eddington-IrrNG}. 

We see however that the error induced by the Eddington approximation remains lower than $5\%$, with the deep temperatures being colder in the analytical model than in the numerical model. Compared to other sources of uncertainty (in particular our assumptions that $\beta$ and $\kappa_2/\kappa_1$ are uniform in the atmosphere), this is an acceptable level of uncertainty. 

\section{Consequences of convection on the overlaying radiative solution}
\label{sec::Convection}
At high-enough optical depth, the deep atmospheres of giant planets and brown dwarfs become convective \citep[\eg][]{Guillot2005}, a consequence of the increase in the opacity with pressure \citep[see][]{Rauscher2012b}. This increase in the opacity in substellar atmospheres is due both to collision-induced absorption by hydrogen molecules increasing with density (above roughly $10^{-3}\rm g\,cm^{-3}$) and eventually to new opacity sources linked to a larger abundance of electrons at temperatures $\sim 2000$\,K and above. Generally, exoplanets and brown dwarfs with low-irradiation levels (\ie such that $T_{\rm irr}\wig{<} T_{\rm int}$) have a convective zone extending all the way from the deep interior to the $\tau\sim 1$ optical depths. This is for example the case of Jupiter, whose atmosphere becomes convective at pressures of order $P\sim 0.3\,\rm{bar}$ -- but with considerable heterogeneity depending on the latitude and longitude on the planet \citep[\eg][]{Magalhaes2002,West2004}. However, in very close-in exoplanets and brown dwarfs, the high stellar irradiation maintains the atmosphere in a very hot state and pushes the radiative/convective transition down to very high optical depths \citep[see][]{Guillot1996,Guillot2005}. 

Numerical models naturally account for these convective zones by imposing a temperature gradient set by convection when a condition such as the Schwarzschild or Ledoux criterion is met. {\tt The temperature profile in the radiative part(s) of the atmosphere is then recalculated taking into account the presence of convective zone(s). The procedure is applied iteratively until a full radiative/convective equilibrium is reached}. {\tt While it is easy to implement the first condition (imposing a convective gradient when necessary) in analytical atmospheric models, it is generally not possible to implement the second one and modify the radiative solution because of the presence of a convective region}. In the specific case of the grey and semi-grey model,~\citet{Robinson2012} recently derived a radiative-convective model that satisfies these two conditions, although it necessitates a small numerical integration. For non-grey thermal opacities, no analytical model solves self-consistently for the convective and the radiative parts of the atmosphere. In the specific case of the model of \citetalias{Parmentier2014a}, the boundary condition of the radiative atmosphere lays in the optically thick layers and a the solution cannot be modified to account for a change in the temperature gradient at deep levels. 

{\tt We want to estimate the error made when the presence of a convective zone is neglected when calculating the temperature profile in the radiative parts of the atmosphere.} We build our analytical radiative/convective model by switching from our radiative solution to the adiabatic solution whenever the convective gradient becomes lower than the radiative one. We compare the resulting analytical solution to the numerical solution in which both the depth of the radiative/convective boundary and the atmospheric temperature profile are converged iteratively. As the presence and depth of a convective zone depends on the exact value of the opacities, we need to specify the Rosseland mean opacity in our model. In this section, in order to facilitate the comparison, we fix the Rosseland mean opacity as a function of pressure to its value in our fiducial model, described in Fig.~\ref{fig::Opacities}. However, all our results will be relative to the depth of the convective zone and thus independent from the exact Rosseland mean opacity function used.

We only consider the case for which the atmosphere transitions from being radiative at high altitudes to being convective at depth (\ie we do not include the possibility of alternating radiative and convective zones). In the convective zone, we assume that the temperature gradient is exactly adiabatic (\ie we do not account for the superadiabatic gradient required to transport the heat flux -- \eg~\citet{Guillot2005}).

\subsection{Non-irradiated grey case}
We first compare the solutions obtained in the non-irradiated grey case. In order to see how the location of the radiative/convective zone influences the solutions, we artificially modify the adiabatic gradient by a factor that varies from $1/4^{th}$ to $4$.

\begin{figure}[htbp]
\includegraphics[width=\linewidth]{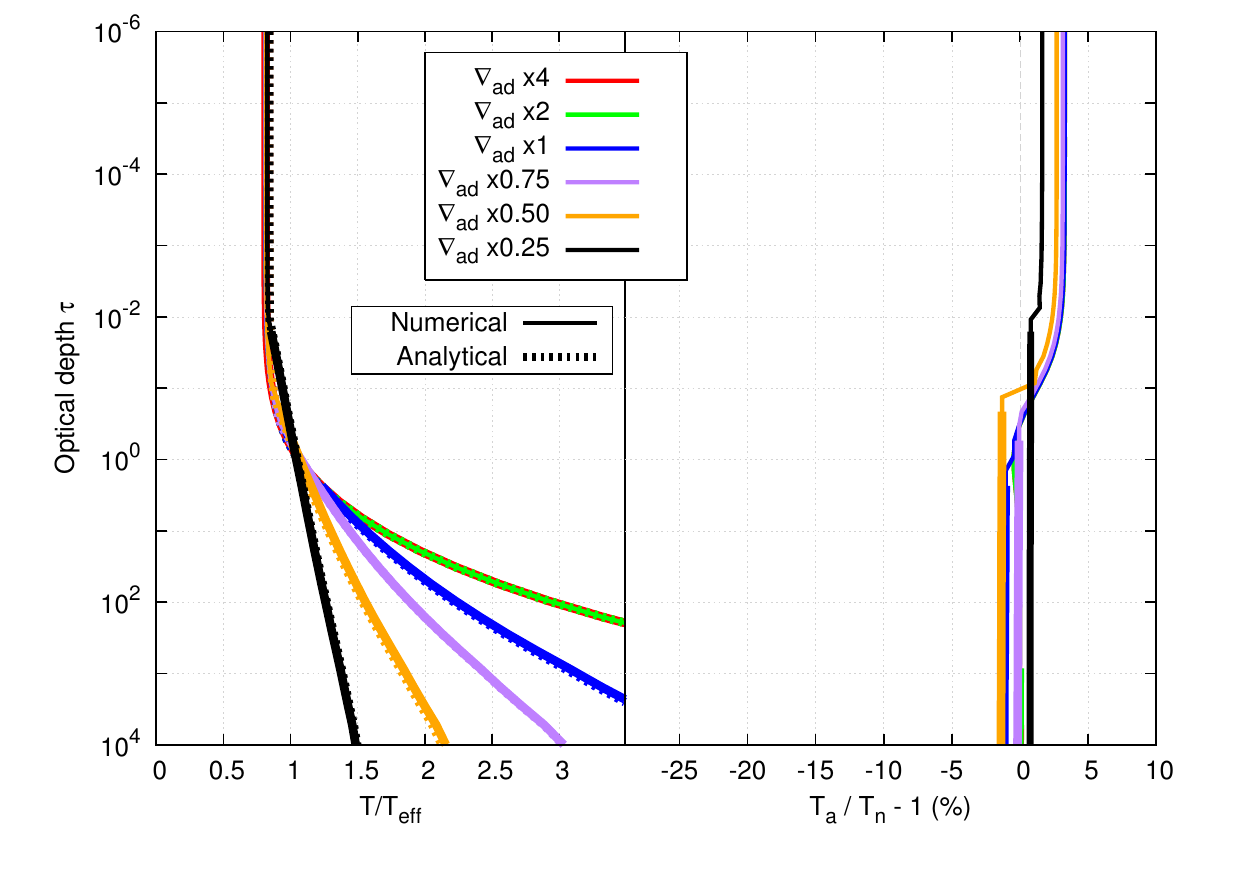}
\caption{Comparison of our numerical and analytical radiative-convective models for different adiabatic gradients in the non-irradiated, grey case. The thin line is the radiative zone and the thick one represents the convective zone. We used $T_{\rm int}=\unit{500}K$ and $g=\unit{25}\meter\per\second\squared$. Note that the cases $\nabla_{\rm ad}\times2$ (green) and $\nabla_{\rm ad}\times4$ (red) are superimposed.}
\label{fig::ConvGreyNoIrr}
\end{figure}

When the radiative/convective transition occurs below optical depth unity (red, green, blue and purple curves in Fig.~\ref{fig::ConvGreyNoIrr}), the difference between the analytical and numerical solutions is unchanged (the corresponding curves are indistinguishable on the right panel) and entirely due to the Eddington approximation as discussed in the previous section. This error is frozen at the radiative/convective boundary and propagates in the convective zone leading to an estimate of the deep temperature profile that is at most 2\% percent off. For a convective zone that crosses the $\tau\approx1$ limit (orange and black curves of Fig.~\ref{fig::ConvGreyNoIrr}), the lower boundary condition used in the analytical radiative model -- that the deep atmosphere reaches the diffusion limit -- is no more valid. The error becomes dependent on the location of the radiative/convective transition (and value of the adiabatic gradient). It however remains of the same order as the one due to the Eddington approximation. This validates models calculating the radiative/convective boundary of the deep convective zone without re-calculating the upper radiative profile. However, the presence of detached convective zones cannot be modeled correctly with this method, and an approach similar to~\citet{Robinson2012} is needed.

\subsection{Non-irradiated non-grey case}

\begin{figure}[htbp]
\includegraphics[width=\linewidth]{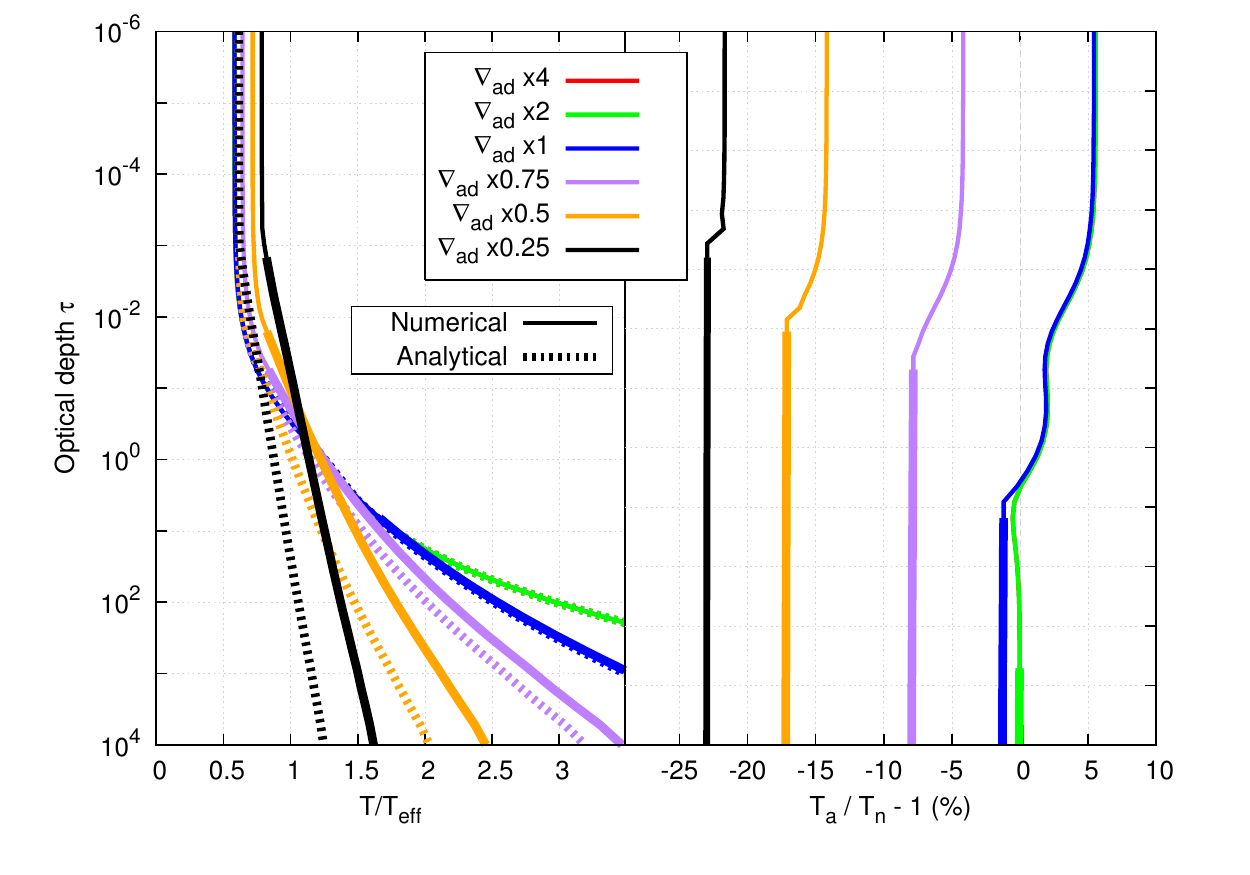}
\caption{Comparison of our numerical and analytical radiative-convective solutions for different adiabatic gradients in the non-irradiated, non-grey case. The thin line is the radiative zone and the thick one represents the convective zone. We used $T_{\rm int}=\unit{500}K$, $g=\unit{25}\meter\per\second\squared$, $\Kb/\Ka=10^{2}$, and $\beta=0.83$. The cases $\nabla_{\rm ad}\times2$ (green) and $\nabla_{\rm ad}\times4$ (red) are superimposed.}
\label{fig::ConvNonGreyNoIrr}
\end{figure}

We now turn, with Fig.~\ref{fig::ConvNonGreyNoIrr}, to the non-irradiated non-grey case, using the fiducial values $\Kb/\Ka=10^{2}$ and $\beta=0.83$. As in the grey case, the errors are dominated by the Eddington approximation as long as the radiative/convective boundary occurs at optical depths larger than unity in the two thermal channels that are considered. The error at low optical depths is larger because the error due to the Eddington approximation is greater in the non-grey case. However, as soon as the convective zone extends to levels of optical depth unity or smaller, the discrepancy between the analytical and numerical solutions increases significantly: the upper atmosphere warms up and our analytical solution is no more a good representation of the radiative atmosphere. This is clearly due to a non-grey effect. In a given spectral interval, the thermal flux present a given pressure is set by the integrated thermal emission of all the atmospheric layers below it in this specific spectral interval. At large optical depth, the emission is thermalized and the thermal flux per wavelength emitted in both spectral channels is the same regardless of the temperature gradient. At optical depth close to unity, the thermal flux in each channel depends on the actual temperature gradient. The analytical solution assumes that the temperature gradient is set by radiation transport everywhere and thus calculates inaccurately the flux emitted in the two spectral channels if convection extends to optical depths smaller than unity. The resulting temperature profile can differ by tens of percentage points from the numerical one. In addition, because the relative error is frozen at the one obtained at the radiative/convective transition, it does not tend towards zero at large optical depths as was the case with the purely radiative solutions. 

Considerable caution should therefore be exerted when switching from radiative to convective gradient without recalculating the radiative solution in the general (non-grey) case. Specifically, when the atmosphere becomes convective at optical depths smaller than unity, the resulting temperature profile may be inaccurate by several tens of percentage points.

\subsection{Irradiated non-grey case}

We now consider the effect of irradiation with our fiducial hot Jupiter atmosphere. As already discussed, the strong irradiation tends to push the radiative/convective zone towards deep levels \citep[see][]{Guillot2005}. This is seen in the profiles of Fig.~\ref{fig::ConvNonGreyNoIrr} which all occur at optical depths $\sim 100$ or deeper, with only a small dependence on the value of the chosen adiabatic gradient. As expected, this suppresses the changes in the temperature profile in the purely radiative atmosphere. The errors are almost independent of the assumed adiabatic gradient and mostly due to the Eddington approximation. For hot Jupiters, and generally for strongly irradiated atmospheres, the presence of a deep convective zone may be accounted for by adopting a purely radiative solution and switching to the convective one when the Schwarzschild criterion is verified. 

Of course, for a smaller irradiation level and/or larger values of the $\Kb/\Ka$ ratio, the presence of a convective zone reaching optical depths closer to unity (in one of the thermal channels at least) will lead to an increase on the error of the calculated temperature profile. We expect this error to be approximately bounded by that of the non-grey, non-irradiated case.

\begin{figure}[htbp]
\includegraphics[width=\linewidth]{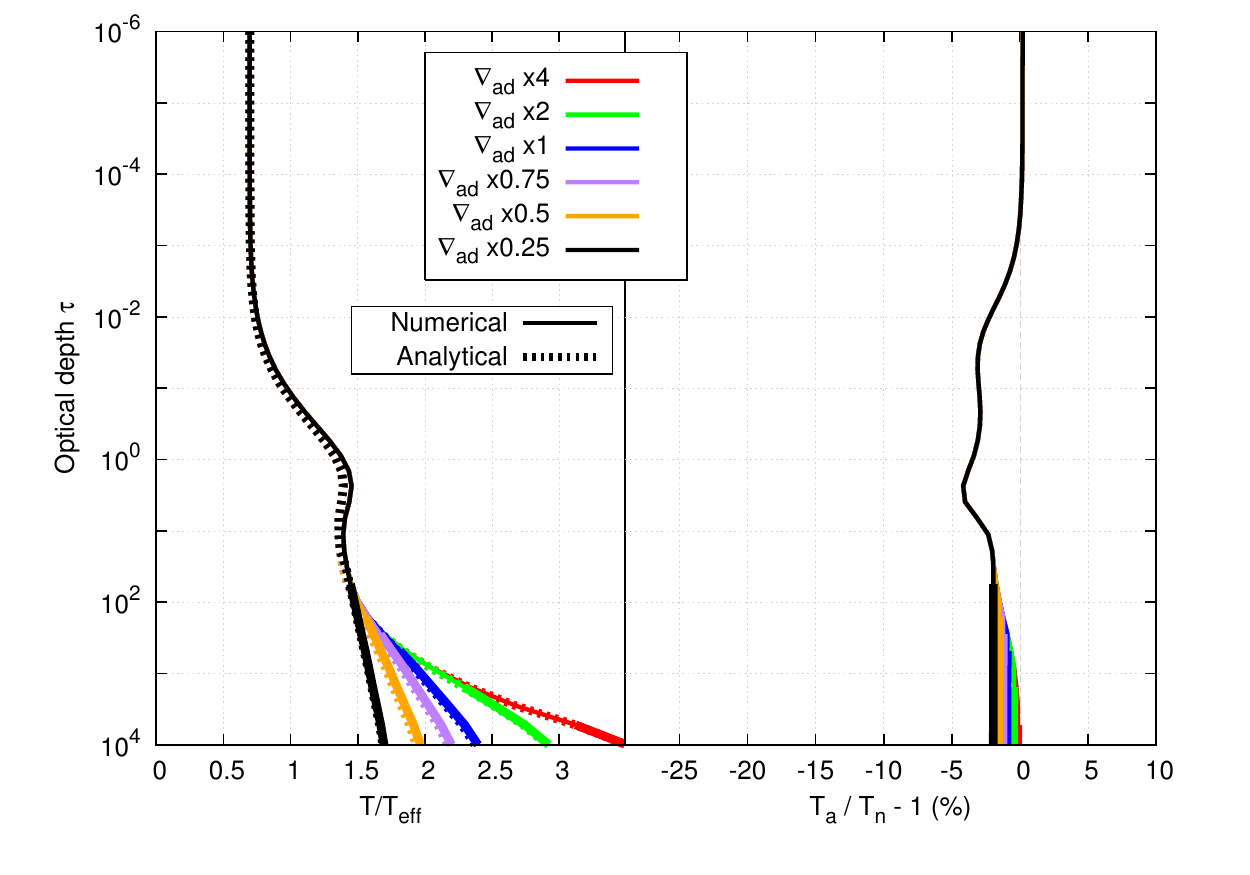}
\caption{Comparison of our numerical and analytical radiative/convective solutions for different adiabatic gradients in the non-grey, irradiated case. The thin line is the radiative zone and the thick one represents the convective zone. We used $\tirr=\unit{1288}K$, $T_{\rm int}=\unit{500}K$, $\Gv=0.25$, $\mu=1/\sqrt{3}$, $\Kb/\Ka=10^{2}$ and $\beta=0.83$.}
\label{fig::ConvNonGreyIrr}
\end{figure}

\section{Modeling the non-grey effects}

Analytical model atmospheres are useful to understand the key physical processes at stake in planetary atmospheres. Unfortunately they cannot take into account the complex variation of the opacities with frequency, temperature and pressure. However, when modeling a specific planet atmosphere with a given chemical composition, the knowledge of the line-by-line opacities should {\tt guide one in a proper} choice of parameters when using the analytical models. In this section we wish to understand what characteristics of the opacities shape the temperature/pressure profile of a planet atmosphere and find a method to derive the simplified opacities of our analytical model from the line-by-line opacities. Ideally, the resulting analytical temperature/pressure profile should be a good approximation of the numerical solution computed with the full frequency, temperature and pressure dependent opacities.

A first approach to determine our coefficients is an \emph{a-posteriori} determination \ie~to choose the coefficients such that the analytical and the numerical profiles match correctly. Although this should give the best results in terms of goodness of the fit, the retrieved coefficients might not be physically realistic and it could be difficult to relate them to the real atmospheric opacities. Another approach is to find \emph{a-priori} values, directly from the opacities. This requires a deep understanding of the opacities and how they shape the temperature profile. A last possibility is to combine the two approaches: using an \emph{a-priori} determination when possible and adjusting the remaining coefficients \emph{a-posteriori} to fit the numerical profile.

\label{sec::Results}

\subsection{A priori determination of the coefficients}
\label{sec::Apriori}
\subsubsection{Visible coefficients}
The visible coefficients control at which depth the stellar flux is absorbed in the atmosphere. When the visible absorption is strong, the stellar flux is absorbed in the upper part of the atmosphere and radiated back to space. At the opposite, when the visible absorption is weak, the incoming irradiation is deposited at depth where the thermal optical depth is large and the deep atmosphere warms up. This is the well-known greenhouse effect. 

When taking into account only one visible band (\ie~$\betav=1$, as in ~\cite{Guillot2010}), a natural choice for the parameter $\Ga$ is the ratio of the mean Rosseland visible opacity (using the Planck function to weight the line-by-line opacities) to the mean Rosseland thermal opacity (using the local Planck function at the stellar effective temperature to weight the line-by-line opacities). Unfortunately, this ratio can vary significantly with height. We find that choosing the ratio at $\tau_{\rm v}=2/3$ (where $\tau_{v}$ is the Rosseland visible optical depth) leads to a correct representation of the absorbed stellar flux and could be used, together with a correct modeling of thermal non-grey effects, to get a first guess of the deep temperature. The part of the stellar flux that heats up the deep atmosphere is the one that propagates down to the $\tau>1$ level. Thus, the opacities that determine the relevant strength of the visible absorption are the lowest visible opacities. The Rosseland mean is a good estimate of the weakest opacities over a given frequency range and is thus a suitable estimate.

However, when a significant portion of the stellar radiation is absorbed in the upper atmosphere of the planet, for exemple when strong visible absorbers such as titanium oxide or sodium are present in the atmosphere, the stellar flux that reaches the $\tau>1$ level depends strongly on the amount of absorption in each spectral channels in the upper atmosphere. The knowledge of $\Gv$ at a given level is not sufficient anymore for a correct estimate of the deep temperature. A more sophisticated model of the visible absorption is then needed.

\begin{figure}[h!]
\includegraphics[width=\linewidth]{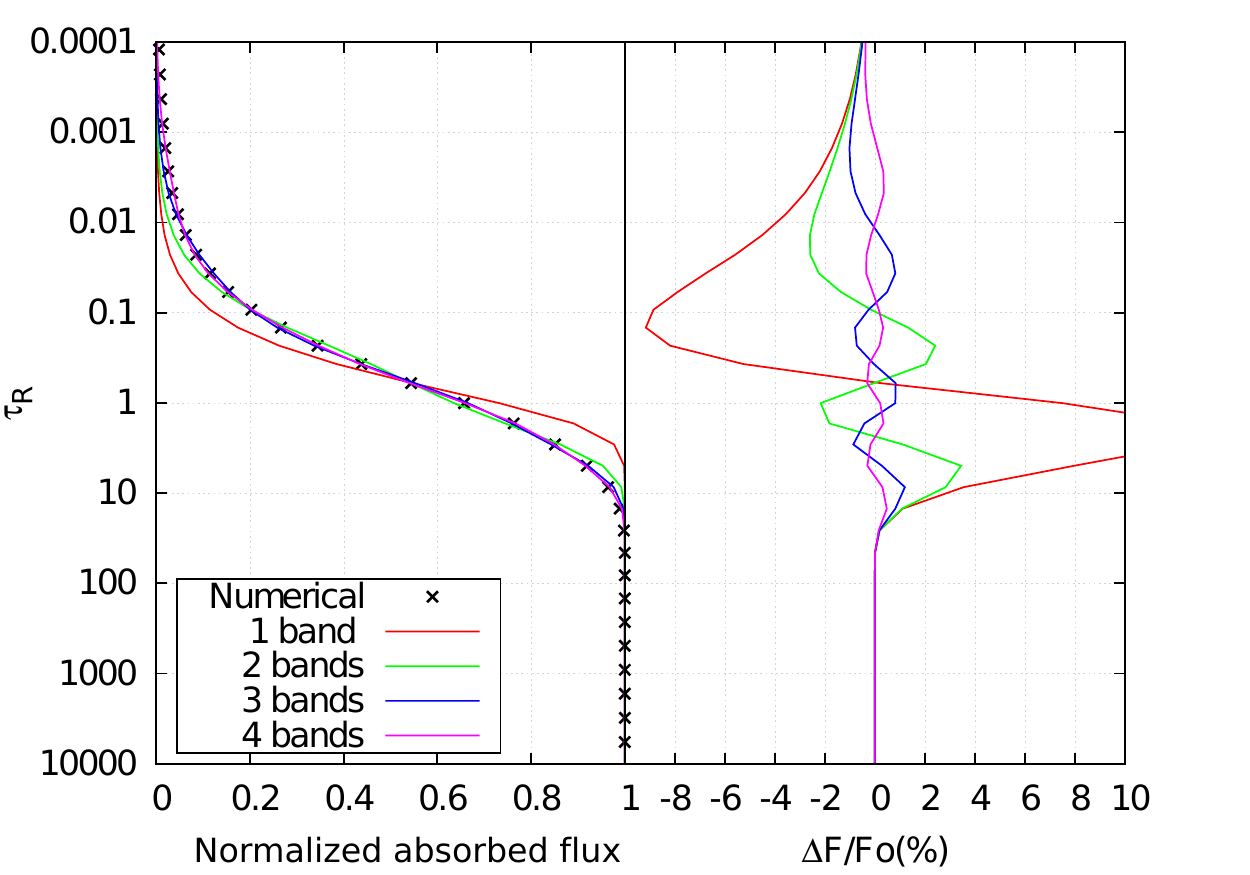}
\caption{Absorbed stellar flux from the numerical model (dots) and from the analytical model (lines) considering respectively 1, 2, 3 or 4 absorption bands in the visible for our fiducial model atmosphere (see Fig.~\ref{fig::Opacities}).}
\label{fig::Absorbed_Flux}
\end{figure}

As shown in~\citetalias{Parmentier2014a}, the moment equations are linear with respect to the absorbed stellar flux. Thus our model can take into account as many spectral bands in the visible as needed with the condition that the different values $\gamma_{\rm{v}\mathit{i}}=\kappa_{\rm{v} \mathit{i}}/{\kappa_{\rm R}}$ in each visible bands are constant through the atmosphere. 
If well chosen, constant non-grey visible opacities can relatively well approximate the absorbed stellar flux at all atmospheric levels. We therefore adopt the following method: using the line-by-line opacities from \citet{Freedman2008} and the actual numerical PT profile, we calculate the total absorbed flux at each layer of the atmosphere. We then adjust the relative contributions of the different visible opacity bands in order to correctly match the absorbed visible flux from the numerical simulation. The stellar flux absorbed by $n$ spectral bands of width $\beta_{\rm v \it{i}}$ is
\begin{equation}
F(\tau)=F_{0}\sum_{i=1}^{n}\beta_{\rm v \it{i}}e^{-\gamma_{\rm{v}\it{i}}\tau/\mu_{*}}\,,
\end{equation}
where the visible bands are homogeneously distributed in frequency (similar to the thermal bands), $F_{0}$ is the total incident stellar flux and the $\beta_{\rm v \it{i}}$ must verify: $\sum_{i}\beta_{\rm v \it{i}}=1$.
{\tt We apply this method using one to four opacity bands. As seen in Fig.~\ref{fig::Absorbed_Flux}, the absorbed flux is poorly represented when considering only one opacity band. When using two, three or four bands, the absorbed flux is described with a $4\%$, $1\%$, and $0.5\%$ accuracy, respectively. In the following we limit ourself to three opacity bands of constant width $\beta_{{\rm v}i}=1/3$ for the visible opacity.}

\begin{figure}[h!]
\includegraphics[width=\linewidth]{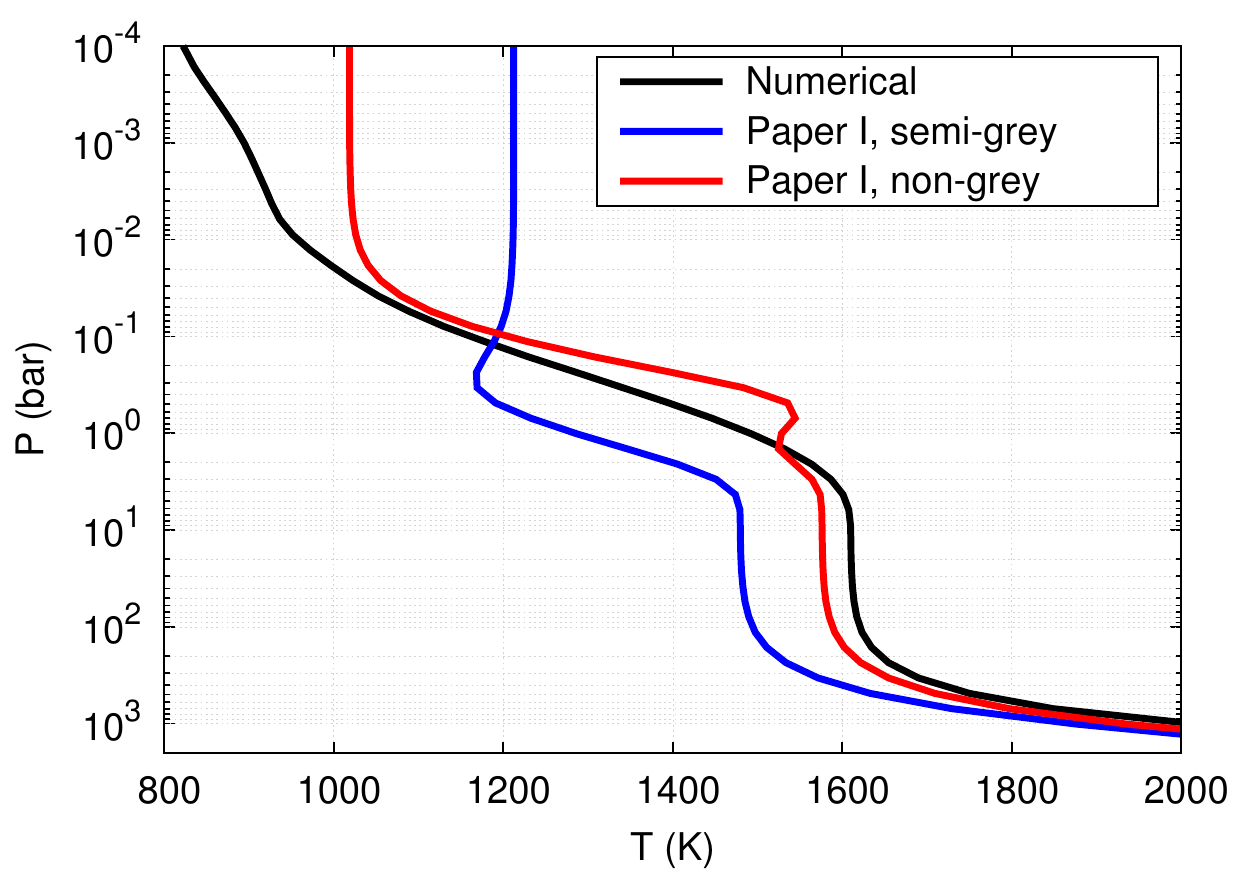}
\caption{{\tt Pressure-temperature profiles calculated using the numerical model and the full set of opacities (black), the semi-grey (blue) and the non-grey (red) analytical. As un Fig.~\ref{fig::Opacities}, $g=\unit{25}\meter\per\second\squared$, $\mu_{*}=1/\sqrt{3}$, $\tint=\unit{100}K$ and $\tmu=\unit{1253}K$. The coefficients used for the analytical models are taken from Table~\ref{table::Models} with $\Gp=1$ for the semi-grey case. The non-grey model is a much better match to the numerical profile than the semi-grey one.}}
\label{fig::Semi-grey}
\end{figure}

Fig.~\ref{fig::Semi-grey} compares the numerical model in black, taking into account all the line-by-line opacities and the semi-grey model (blue line) where the visible opacities are adjusted in order to have the same absorbed flux as in the numerical model but where the thermal opacities remain grey. The semi-grey model, \emph{even though it models correctly the absorbed flux as a function of depth}, lays far from the numerical solution. Clearly, non-grey thermal opacities are needed.

\subsubsection{Thermal coefficients}
\begin{figure}
\includegraphics[width=\linewidth]{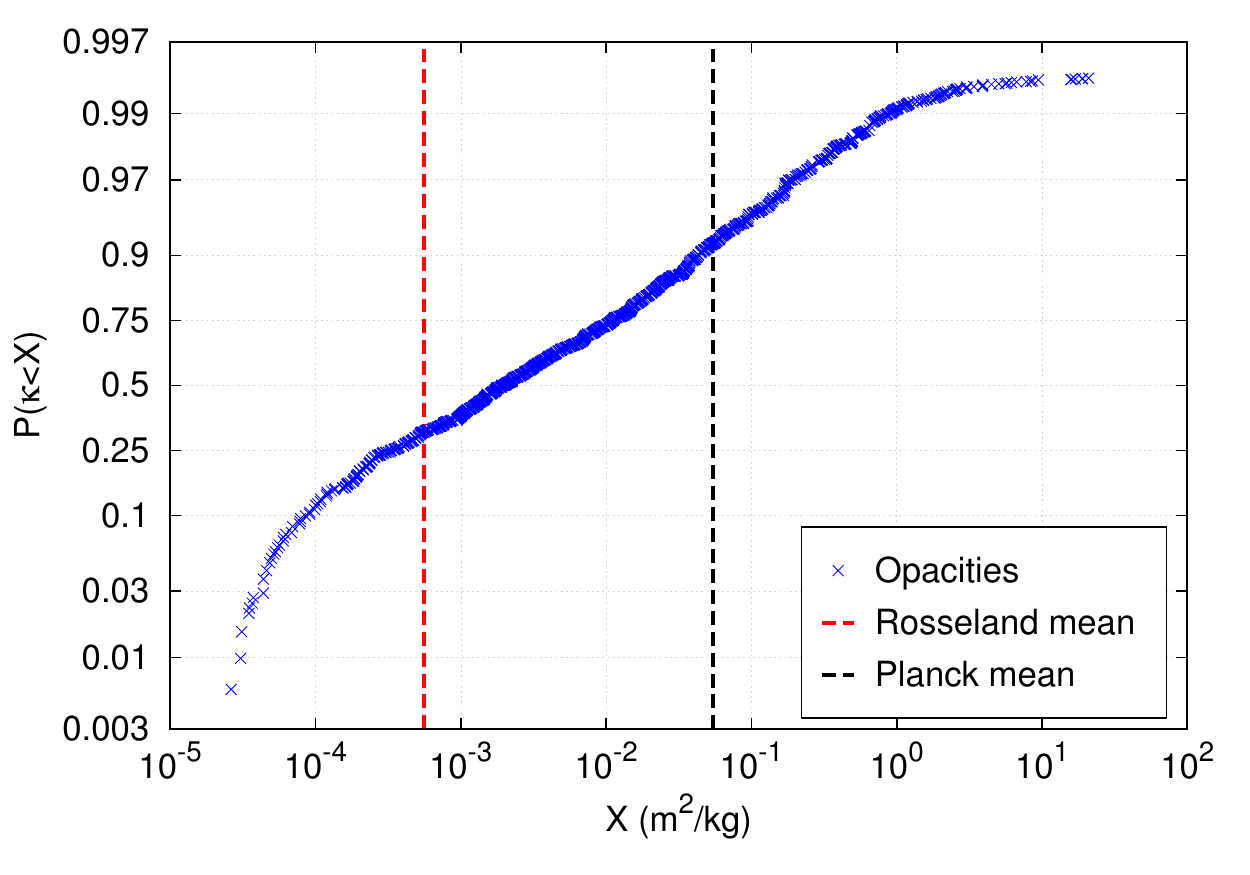}
\caption{Cumulative distribution function of the opacities at $P=\unit{0.85}\bbar$ and $T=\unit{1464}K$, corresponding to the $\tau=2/3$ level of an atmosphere with $\tmu=\unit{1253}K$ and a gravity $g=\unit{25}\meter\per\second\squared$. The Y axis represents the fraction of frequency where the monochromatic opacities are lower than the corresponding $\kappa_{0}$ of the X-axis. The red line shows the value of the Rosseland mean opacity and the black line the Planck mean opacity. $25\%$ of the frequency range have monochromatic opacities smaller than the Rosseland mean opacity whereas $90\%$ have monochromatic opacities smaller than the Planck mean opacity.}
\label{fig::Repartition} 
\end{figure}
The thermal coefficients describe how well the atmosphere is able to retain its energy. As explained qualitatively by~\citet{Pierrehumbert2010} and quantitatively in~\citetalias{Parmentier2014a}, the presence of non-grey thermal opacities can strongly affect the temperature profile of the planet. Because it is tied to the emission and absorption of the thermal flux, only the opacity variations that have an extent smaller or comparable to the local Planck function can contribute to the non-grey effects. The cumulative distribution function of the opacities in the frequency range covered by the local Planck function should thus contain enough information to constrain the non-grey effects. As a grey atmosphere cools down principally by emission from the $\tau=2/3$ level, the {\tt thermal opacities} at this level should determine the strength of the non-grey effects.

We plot in Fig.~\ref{fig::Repartition} the cumulative distribution function of the opacities at the $\tau=2/3$ level for our fiducial model atmosphere. It represents the relative spectral width over which the opacities are lower than a given opacity $\kappa_{0}$ as a function of $\kappa_{0}$. The opacities cover a wide range of value (6 orders of magnitude in the specific example shown in Fig.~\ref{fig::Repartition}). Our analytical model can describe the non-grey thermal opacities with only two parameters: the ratio of the Planck mean opacity to the Rosseland mean opacity, $\Gp$ and the relative size of the two bands, $\beta$. Unlike in the visible case, the thermal effects are local effects that do not depend on the behavior of the rest of the atmosphere. The value of $\Gp$ can hence be calculated as a function of pressure and temperature from tables available in the community \citep[\eg][]{Freedman2008}. 

The parameter $\beta$ describes the relative amount of the opacities which are in the first band compared to the second band. The Rosseland mean opacity is determined by the smallest values of the opacities, which is the second band opacity in our model. We decide to use as $\beta$ the fraction of the opacities in the spectral range covered by the local Planck function that are higher than the Rosseland mean opacity. This can be derived directly from the cumulative distribution function of the opacities plotted in Fig.~\ref{fig::Repartition}. In the specific example of Fig.~\ref{fig::Repartition}, $25\%$ of the opacities lay below the Rosseland mean opacity hence $\beta=0.75$.

\begin{figure*}[ht!]
\center
\includegraphics[width=0.82\linewidth, trim=0cm 0cm 0cm 0cm, clip=true]{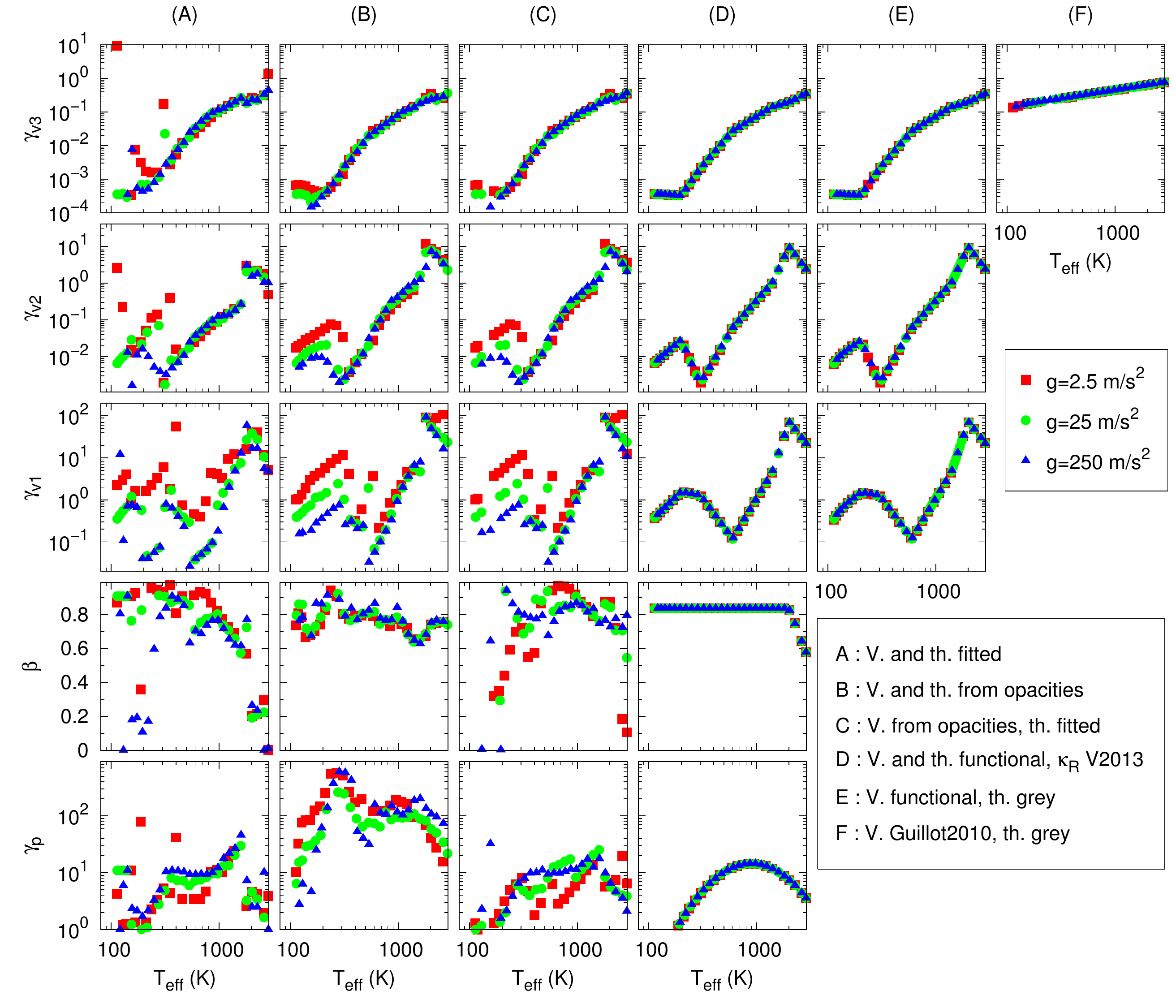}
\includegraphics[width=0.82\linewidth, trim=0cm 0cm 0cm 0cm, clip=true]{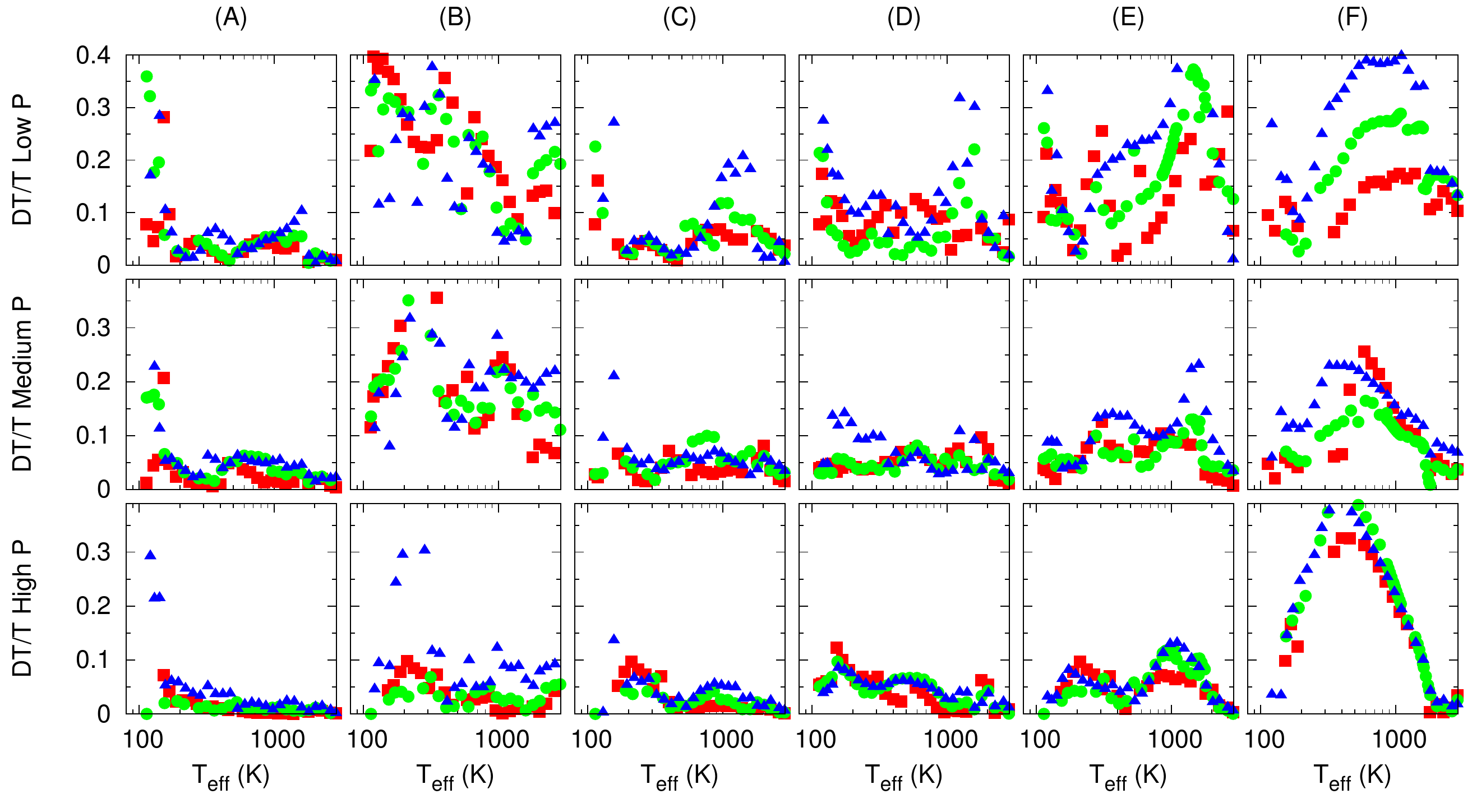}
\caption{
\emph{Top panel}: coefficients $\Gvc$, $\Gvb$, $\Gva$, $\beta$, and $\Gp$ obtained for the six different models described in Sect.~\ref{sec::Application} as a function of the irradiation temperature for planets of solar composition with different gravities and an internal temperature of $100\kelvin$.\\
\emph{Bottom panel}: Mean relative difference between the numerical and the analytical model for the six different models described in Sect.~\ref{sec::Application}. The first line is the mean difference for $\unit{10^{-4}}\bbar<P<\unit{10^{-2}}\bbar$, the second one for $\unit{10^{-2}}\bbar<P<\unit{10^{0}}\bbar$ and the third one for $\unit{10^{0}}\bbar<P<\unit{10^{2}}\bbar$. In terms of Rosseland optical depth, the low pressure zone corresponds to the optically thin part of the atmosphere with $10^{-8}(25\,{\rm m s^{-2}}/g)\lesssim\tau_{\rm R}\lesssim10^{-2}(25\,{\rm m s^{-2}}/g)$. The medium pressure zone corresponds to the transition from optically thin to optically thick with $10^{-4}(25\,{\rm m s^{-2}}/g)\lesssim\tau_{\rm R}\lesssim10(25\,{\rm m s^{-2}}/g)$. The high pressure zone corresponds to the optically thick part of the atmosphere with $(25\,{\rm m s^{-2}}/g)\lesssim\tau_{\rm R}\lesssim10^{4}(25\,{\rm m s^{-2}}/g)$.}
\label{fig::Results} 
\end{figure*}

\begin{center}
\begin{table*}

\caption{Functional form of the coefficients of the analytical model of~\citetalias{Parmentier2014a} valid for solar composition atmospheres. We use $X=\log_{10}(\teff)$}
\centering
\ra{1.7} 
\begin{tabular}{>{\centering}m{1.5cm} >{\centering} m{1.9cm} >{\centering} m{1.9cm} >{\centering} m{1.9cm} >{\centering} m{1.9cm} >{\centering} m{1.9cm}  >{\centering} m{1.9cm}  >{\centering} m{1.9cm}}

\toprule 
Coefficient & Expression & $\teff<200\,\K$ & $200\,\K<\teff<300\,\K$ & $300\,\K<\teff<600\,\K$ & $600\,\K<\teff<1400\,\K$& $1400\,\K<\teff<2000\,\K$ & $\teff>2000\,\K$\\ 
\midrule
\multirow{2}{*}{$\log_{10}(\Gvc)$}&\multirow{2}{*}{$a+bX$}&$a=-3.03$&$a=-13.87$&$a=-11.95$&$a=-6.97$&$a=-3.65$&$a=-6.02$\\
&&$b=-0.2$&$b=4.51$&$b=3.74$&$b=1.94$&$b=0.89$&$b=1.61$\\\hline
\multirow{2}{*}{$\log_{10}(\Gvb)$}&\multirow{2}{*}{$a+bX$}&$a=-7.37$&$a=13.99$&$a=-15.18$&$a=-10.41$&$a=-19.95$&$a=13.56$\\
&&$b=2.53$&$b=-6.75$&$b=5.02$&$b=3.31$&$b=6.34$&$b=-3.81$\\\hline
\multirow{2}{*}{$\log_{10}(\Gva)$}&\multirow{2}{*}{$a+bX$}&$a=-5.51$&$a=1.23$&$a=8.65$&$a=-12.96$&$a=-23.75$&$a=12.65$\\
&&$b=2.48$&$b=-0.45$&$b=-3.45$&$b=4.33$&$b=7.76$&$b=-3.27$\\\hline
\multirow{2}{*}{$\beta$}&\multirow{2}{*}{$a+bX$}&$a=0.84$&$a=0.84$&$a=0.84$&$a=0.84$&$a=0.84$&$a=6.21$\\
&&$b=0$&$b=0$&$b=0$&$b=0$&$b=0$&$b=-1.63$\\\hline
$\log_{10}(\Gp)$&$aX^{2}+bX+c$&\multicolumn{6}{c}{\hfill$a=-2.36$ \hfill $b=13.92$\hfill$c=-19.38$\hfill}\\\hline
\bottomrule 
\label{table::Models}

\end{tabular}
 
 \end{table*}
\end{center}

\begin{SCfigure*}
  \includegraphics[width=12cm,clip]{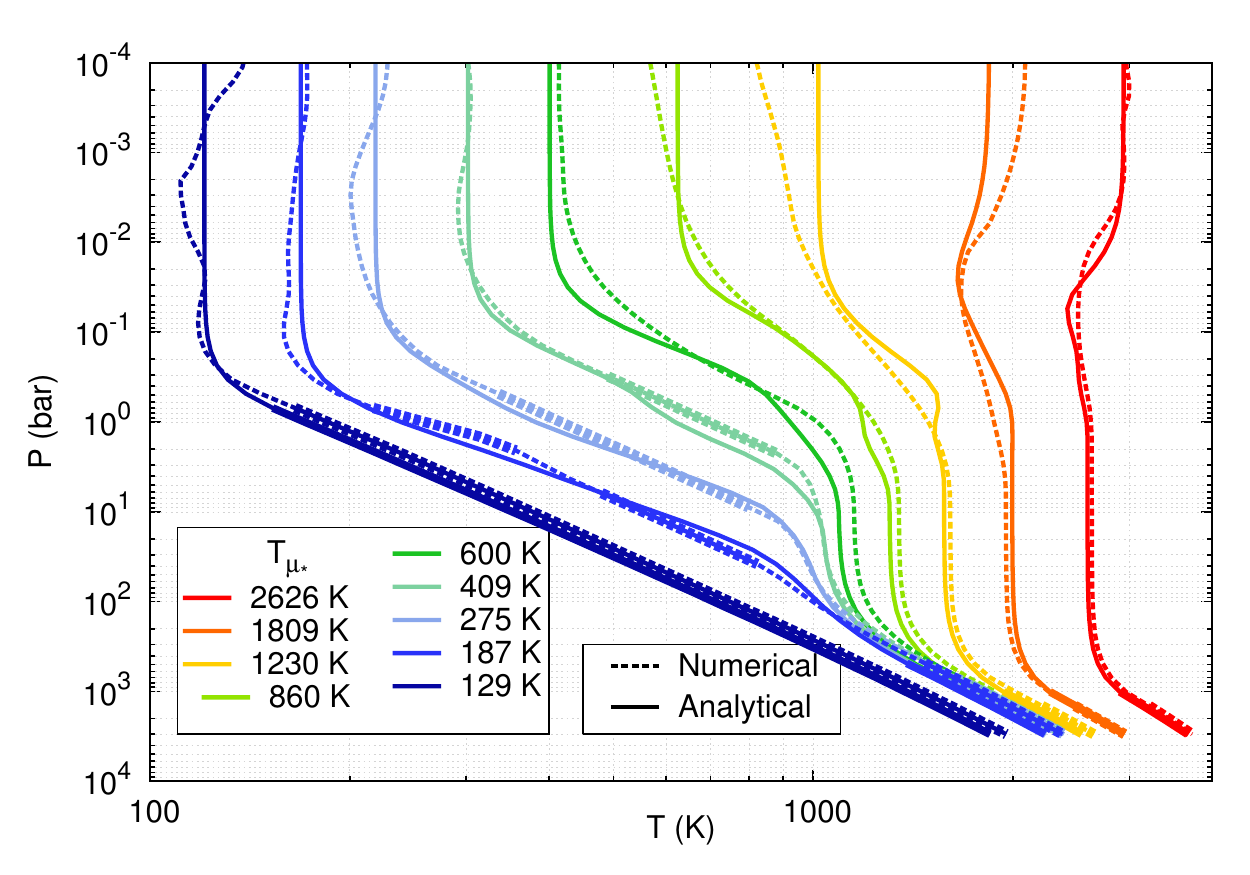}
\caption{Comparison between the numerical solutions (dashed lines) and the analytical solutions of model D (using the functional form of the coefficients given in Table~\ref{table::Models}) over a wide range of irradiation temperatures for a giant planet of solar composition orbiting a sun-like star. Here we used $g=\unit{25}\meter\per\second\squared$, $\tint=\unit{100}K$, and $\mu_{*}=1/\sqrt{3}$. Thick and thin lines represent convective and radiative zones, respectively.}
\label{fig::AllProfiles}
\end{SCfigure*}

\subsection{Application/Different models}
\label{sec::Application}

In order to test the goodness of our analytical model and derive reasonable estimates of the coefficients, we use the EGP numerical code to build a grid of atmospheric radiative/convective models for giant planets with a solar composition atmosphere, three different gravity ($2.5$, $25$, and $\unit{250}\meter\per\second\squared$), and an internal temperature of $T_{\rm int}=\unit{100}K$. We consider the case of a planet orbiting a sun-like star at various distances corresponding to irradiation temperatures from $\unit{100}K$ to $\unit{3000}K$. All the profiles were calculated using $\mu_{*}=1/\sqrt{3}$ and therefore represent global or dayside averaged temperature profiles (see Sect.~\ref{sec::Setting} for more details).
Figure~\ref{fig::Results} shows different models obtained for different estimates of our coefficients (top panel) and a comparison between the numerical profiles and the resulting analytical profiles (bottom panel). In all models but model D, we use as Rosseland mean opacity the one calculated by the numerical model directly from the line-by-line opacities. In model D, we use the functional fit of the Rosseland mean opacities of~\citet{Freedman2008} provided by~\citet{Valencia2013}. Similarly, model D uses a functional form for the convective gradient when convective adjustment is necessary (see Sect.~\ref{sec::ResultsConv}. Model D is therefore a fully analytical model that can be downloaded and implemented by the community. We now describe the different models.

{\it Model A:}
In this model, we adjust all our coefficients \emph{a-posteriori} in order to have the best match to the numerical profiles. It leads to temperature profiles in agreement within $5\%$ with the numerical ones. It shows that our analytical model can represent a large variety of atmospheric temperature profiles and goes beyond the limitation of previous semi-grey models~\citep{Parmentier2013}. However, the spread of the retrieved value of the coefficients makes it difficult to derive a trustable functional form and a better approach is needed in order to get a fully analytical model.

{\it Model B:}
{\tt Here we use the methods of Sect.~\ref{sec::Apriori} to determine \emph{a-priori} the various coefficients. 
 The visible coefficients, $\Gva$, $\Gvb$, and $\Gvc$ are determined respectively by the first, the second, and the third thirds of the incoming stellar energy that are absorbed by the atmosphere as the stellar irradiation propagates downward. Thus $\Gva$ is determined by the highest values of the visible opacities, $\Gvb$ by the median ones and $\Gvc$ by the lowest ones. All of them are also proportional to the inverse of the thermal Rosseland mean opacities. They exhibit four different types of behavior as a function of the effective temperature. Those behaviors reflect changes in chemical composition with the irradiation temperature (a plot of the line-by-line opacities for the four different regimes is shown in appendix) :

\begin{itemize}
\item For $\teff < 250\,{\rm K}$, the visible opacities are dominated by Rayleigh scattering and exhibit a slight dependence with gravity. As the effective temperature increases, more absorbers are present in gaseous form and the visible opacities increase.

\item For $250\,{\rm K}<\teff<600\,{\rm K}$, the visible opacities are dominated by the sodium lines at great depth, where the profile is warm enough to have sodium in gaseous state, whereas it is dominated by much weaker lines in the upper atmosphere. As $\teff$ increases, the lines broaden, increasing the lowest values of the opacities and decreasing the highest values of the opacities. As a consequence, $\Gvb$ and $\Gvc$ increase with $\teff$ whereas $\Gva$ decreases.

\item  For $600\,{\rm K}<\teff<1700\,{\rm K}$ the sodium and potassium become the main gaseous absorbers in the upper atmosphere, leading to a strong visible absorption and an increase in the visible opacities with $\teff$. 

\item For $\teff>1700\,{\rm K}$ titanium and vanadium oxides become the main gaseous absorbers in the upper atmosphere, creating a substantial increase in the visible opacities. This is reflected by the sudden increase in $\Gva$ and $\Gvb$.

\end{itemize}

The thermal coefficients have a smoother behavior with $\teff$. $\beta$ is rather constant and equal to $\approx0.8$. This high value of $\beta$ can be interpreted as a predominance of the molecular bands (\ie~the water and methane bands) to the atomic lines in the thermal opacities. Around $\teff=300\,\kelvin$, $\beta$ reaches values even closer to $1$. At these temperatures, the Planck function at the atmospheric levels of $\tau\approx2/3$ overlaps with the $5\mu\meter$ window in the opacities which is consistent with large values of $\beta$.
The variations of $\Gp$ with $\teff$ have a saddle-like shape with two maxima at $300\,\kelvin$ and $1500\,\kelvin$. At low temperatures, the Planck function of the atmosphere shifts towards large wavelengths ($>10\mu\meter$) for which the opacities are almost constant, leading to small values of $\Gp$. At very high $\teff$, the Planck function of the atmosphere shifts toward smaller wavelengths ($<1\,\mu\meter$) for which the TiO broad-band absorption significantly flattens the opacities, leading to small values of $\Gp$. In between, when the atmospheric Planck function is between $1$ and $10\mu\meter$, the opacities are dominated by the water and methane bands, which raises the value of $\Gp$ up to $\approx100$.

Although this model gives a correct estimate of the profile at high pressure, it leads to errors of $\approx40\%$ at medium and low pressure. Given that the coefficients were all {\tt determined} \emph{a-priori}, reaching a $40\%$ accuracy can be a fair, first guess of the temperature profile. This method could be extended to planets with very different opacities without going through the whole numerical integration of the radiative transfer equation. However, as proven by model A, a much better accuracy can be obtained by the analytical profile and a mixed method with some coefficients derived \emph{a-priori} and others \emph{a-posteriori} can be a good compromise. 

{\it Model C:}
In this model we use a mixed method to derive the coefficients of the analytical model, with some of them being derived \emph{a-priori} and some of them \emph{a-posteriori}. The method to determine the visible coefficients seems robust, as it can give the correct absorbed flux as a function of optical depth in the atmosphere with a $1\%$ accuracy. The method to determine the thermal coefficients is more subject to caution as it is unclear whether the value of $\Gp$ in our analytical model should correspond to the value of $\Gp$ derived from the real opacities. Moreover, our criteria to choose $\beta$ (the fraction of the opacities that are higher than the Rosseland mean opacity) is \emph{ad-hoc} and does not rely on strong physical arguments. At last, there is no strong argument to choose the depth at which those coefficients are calculated. We thus decided to obtain the visible coefficients from the a-priori solution and to fit the thermal ones by adjusting the analytical profile to the numerical profile. The resulting analytical solutions lead to an estimate of the temperature profile that always differs by less than $10\%$ from the numerical solution. 

Compared to model B, only the thermal coefficients are changed in model C. Below $200\,\K$, $\Gp$ is small and $\beta$ is not well defined. For $200\,\K<\teff<2000\,\K$, $\beta$ is roughly constant with $\teff$ with values of approximately $0.8$, in agreement with the a-priori determination in model B. For $\teff>2000\,\K$, $\beta$ decreases slightly with $\teff$, showing that non-grey effects become more important in the upper atmosphere than in the deep atmosphere.

$\Gp$ keeps the same dependency with $\teff$ than in model B but is one order of magnitude smaller. The high values of $\Gp$ derived in model B seem relevant to understand the deep atmospheric structure but are too high to represent the upper atmosphere. Large values of $\Gp$ lead to a stronger cooling of the upper atmosphere and in temperatures cooler than expected from the numerical model.  

{\it Model D:}
In order to produce a fully analytical model we fit a functional form to the coefficients derived in model C as a function of $\teff$. Following the different regimes that we described in model B, we fit different affine functions to the visible coefficients. {\tt Although the transitions at $\approx 250$ and $\approx 1800\,{\rm K}$ are discontinuous, we decided to provide a continuous fit by introducing two different transition zones for $200\,{\rm K}<\teff<300{\rm K}$ and $1400\,{\rm K}<\teff<2000\,{\rm K}$. Ensuring continuity in the fit is important to increase the numerical convergence of numerical models where this fit can be used as a boundary condition (\eg internal structure and evolution models of giant planets).} The thermal coefficients having a much smoother variation with the irradiation temperature, {\tt We use a constant value $\beta=0.8$ with an affine function for $\teff>2000\,\K$ for the parameter $\beta$ } and we use a $2^{nd}$ order polynomial to represent $\Gp$ over the whole temperature range. The functional form of the coefficients are presented in Table~\ref{table::Models}. 

The resulting model matches the numerical profiles over a wide range of irradiation temperatures and planet gravity with an accuracy always better than $10\%$ for pressures ranging from $100\,\bbar$ to $10^{-2}\,\bbar$ and generally better than $10\%$ for pressures ranging from $10^{-2}\,\bbar$ to $10^{-4}\bbar$ (see Fig.~\ref{fig::AllProfiles} and column D of Fig.~\ref{fig::Results}). The fit is slightly worst at low $\teff$ where we do not model the gravity dependance of the visible coefficients and at the transition zone $1400\,\K<\teff<2000\,\K$ where we model as continuous a discontinuous transition.
}
Whereas for previous models the Rosseland mean opacities used in the analytical solution were calculated by the numerical model directly from the line-by-line opacities, Model D uses the fit of the~\citet{Freedman2008} Rosseland mean opacities provided by ~\citet{Valencia2013}. Moreover, model D includes a functional form of the convective gradient (see Sect.~\ref{sec::ResultsConv} for more details). This makes model D a self-consistent fully analytical model.

{\it Model E:}
Here, the importance of non-grey effects are tested. This model has grey thermal opacities (\ie~$\Gp=1$) but uses the functional form derived in model D for the visible coefficients. Therefore, model E is a good representation of the absorption of the stellar irradiation by the atmosphere but lacks the non-grey effects. At medium and large pressures, model E gives a reasonable estimate of the temperature profile, although worst than model D. At low pressures model E provides a very bad estimate of the temperature, with discrepancies of up to a factor of two with the numerical model. The non-grey absorption of the stellar irradiation therefore cannot by itself explain the temperature structure in planetary atmospheres. Non-grey thermal effects, such as the ones considered in model D, are necessary.

\begin{figure}
  \includegraphics[width=\linewidth]{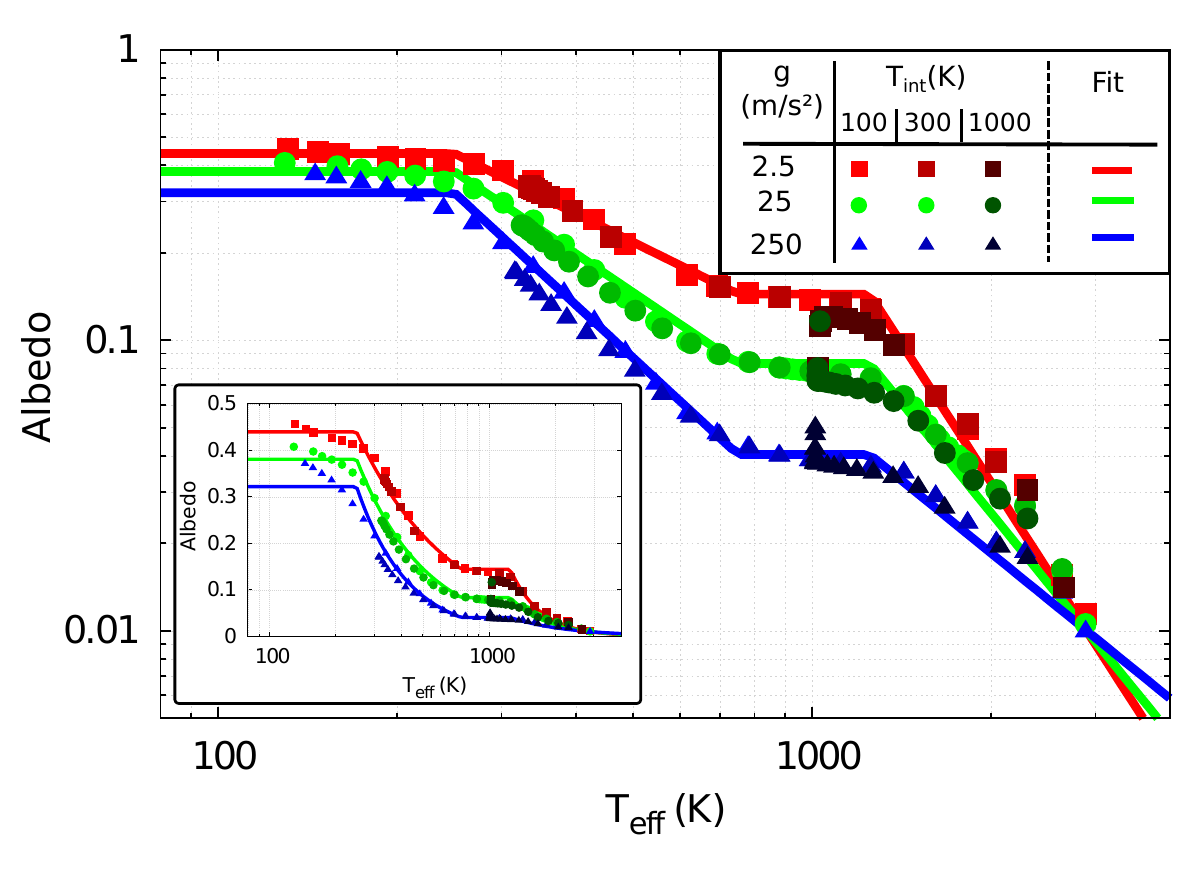}
\caption{{\tt Plane-parallel albedos for solar-composition, clear-sky atmospheres with different values of gravity and internal temperature, numerical model (dots) and analytical fit described in Table~\ref{table::Albedo} (lines). In the inset, the albedo plotted in linear scale. Here we approximate the Bond albedo of the planet by the plane-parallel albedo.}}
\label{fig::Albedo}
\end{figure}

{\it Model F:}
In this model, a comparison with the previous estimate of~\citet{Guillot2010} is done. We use grey thermal opacities and the visible coefficients provided by~\citet{Guillot2010}. Those coefficients were derived in order to match the deep temperature of highly irradiated planets, which it does well. However, at low pressures model F always fails to represent the numerical temperature profile and, at temperatures smaller than $2000\,\K$ the discrepancy can reach $40\%$ at all atmospheric pressures.
{\tt
\subsection{Albedo}
\label{sec::Albedos}
Although our analytical model has been derived for a purely absorbing atmosphere, we indirectly take into account scattering via the value of the Bond albedo. Our numerical model calculates the reflectivity $A_{\mu_{*}}$ of a plane-parallel atmosphere irradiated with an angle $\mu_{*}=1/\sqrt{3}$. Exact calculation of the Bond albedo of a planet involves an integration of the radiative transfer equation for numerous angles of the incident stellar beam and is beyond the scope of this paper~\citep[\eg][]{Cahoy2010}. We hereafter approximate the Bond albedo of the planet by the plane-parallel albedo calculated for $\mu_{*}=1/\sqrt{3}$. This is similar to the so-called \emph{isotropic approximation} for the mean temperature profile~\citep[\eg][]{Guillot2010}. Although the approximation is a strong one, we expect $A_{\rm B}$ and $A_{\mu_{*}}$ to have similar values and to follow the same trends.

As shown in Fig.~\ref{fig::Albedo}, hotter planets have lower albedos. We consider cloudless atmospheres, neglecting all types of clouds that might be present in this temperature range. The derived albedos are therefore lower bounds as any additional diffusion by clouds should increase the planet albedo. Here, the albedo is set by the competition between Rayleigh scattering and atomic/molecular absorption. Hotter planets have larger abundances of gaseous absorbers in their atmospheres and have therefore a lower albedo. At high effective temperatures ($\teff\gtrsim1250K$), titanium and vanadium oxides broadband absorption together with collision-induced absorption dominate the optical opacities, leading to Bond albedos lower than $0.1$. As the effective temperature decreases, TiO and VO become less abundant and the albedo increases. From $750K\lesssim\teff\lesssim1250K$, the visible opacities are dominated by the absorption lines of sodium and potassium, leading to a plateau in the value of the Bond albedo. At lower effective temperatures, the sodium and the potassium slowly disappear from the atmosphere and, at $\teff\lesssim250K$, the albedo reaches its maximum value. 

We also see in Fig.~\ref{fig::Albedo} that planets with higher gravities have lower albedos. This trend is particularly important around $\teff\approx1000\,K$. At these effective temperatures, absorption in the atmosphere is dominated by the pressure-broaden sodium and potassium lines. The pressure of the layer where the photons at a given wavelength are deposited is proportional to the gravity of the planet. For higher gravity planets the stellar photons reach deeper layers where the alkali lines are broader and the absorption higher. Since Rayleigh scattering is independent of pressure, planets with higher gravities absorb more efficiently the stellar irradiation and have lower albedos. 

Interestingly, the albedo is very insensitive to the internal temperature of the planet. For a given effective temperature, changing the internal temperature does not change the temperature profile enough to modify the chemical equilibrium of the atmosphere and change its composition.

Our albedos agree well with the calculations of~\citet{Marley1999} for $\teff=200K$. For $\teff=1000K$ our albedos are much smaller than the ones of~\citet{Marley1999}. As shown in~\citet{Sudarsky2000} alkali metals, not considered in the previous study increase significantly the absorption at those effective temperatures and explain the discrepancy.

In Table~\ref{table::Albedo} we provide a fit of the albedos for different effective temperature and gravities. This fit can be used together with Model D of Sect.~\ref{sec::Application} to obtain a self-consistent model of irradiated planet atmospheres.
}

\begin{center}
\begin{table*}

\caption{Functional form of the planet plane-parallel albedo $\log_{10}(A_{\mu_{*}})=a+bX$ where $X=\log_{10}(\teffo)$ and $g$ is the gravity of the planet in $\rm{m/s^{2}}$. $A_{\mu_{*}}$ can be used as an approximation for the Bond albedo $A_{\rm B}$ of the planet.}
\centering
\ra{1.7} 
\begin{tabular}{>{\centering}m{3cm} >{\centering} m{5cm} >{\centering} m{5cm} }

\toprule 
Temperature range & $a$ & $b$\\ 
\midrule
$\teffo<250\,\K$&$-0.335g^{0.070}$& $0$\\
$250\,\K<\teffo<750\,\K$&$-0.335g^{0.070}+2.149g^{0.135}$& $-0.896g^{0.135}$\\
$750\,\K<\teffo<1250\,\K$&$-0.335g^{0.070}-0.428g^{0.135}$& $0$\\
$\teffo>1250\,\K$&$16.947-3.174g^{0.070}-4.051g^{0.135}$& $-5.472+0.917g^{0.070}+1.170g^{0.135}$\\
\bottomrule 
\label{table::Albedo}

\end{tabular}
 
 \end{table*}
\end{center}

{\tt
\subsection{Radiative/convective model}
\label{sec::ResultsConv}
As discussed in Sect.~\ref{sec::Convection}, the deep atmosphere of substellar objects is convective and the temperature-pressure profile can be determined by integrating the adiabatic gradient :
\begin{equation}
\nabla_{\rm ad}=\left(\frac{\partial\log{T}}{\partial\log{P}}\right)_{S}\,.
\end{equation}
This gradient is an intrinsic property of the fluid and can be derived from the equation of state. Based on a fit of the \mbox{\citet{Saumon1995}} equation of state at high pressures, we use the functional form :
\begin{equation}
\nabla_{\rm ad}\approx0.32-0.1\left(\frac{T}{3000\,\K}\right)\,.
\label{eq::ConvGrad}
\end{equation}
Our radiative/convective model consist of a convective zone from the bottom of the model up to the radiative/convective boundary at a pressure $P_{\rm R/C}$. For $P<P_{\rm R/C}$ the analytical model with the coefficients of model D is used. To choose $P_{\rm R/C}$ we compare the radiative gradient and the convective gradient given by Eq.~\eqref{eq::ConvGrad} from the bottom to the top of the model. As long as the convective gradient is smaller than the radiative gradient the atmosphere is convective. Whenever the radiative gradient becomes smaller than the convective one the radiative/convective boundary is reached. 

Although this method should provide a good estimate of $P_{\rm R/C}$, it fails at low effective temperatures when used with model D. At low effective temperatures, a radiative zone squeezed between two convective zones develops around $P\approx 100\,\bbar$. In the numerical model, the two convective zones merge and the radiative zone disappears for $\teff<200\,\K$. In the analytical model, however, the radiative zone remains and our estimate of $P_{\rm R/C}$ is biased. This discrepancy is due to the accuracy of the fit of the Rosseland optical depth provided by~\citet{Valencia2013}. At $P\approx100\,\bbar$ and $T\approx1000\,\K$ the fit systematically underestimate the Rosseland mean opacities by $\approx30-50\%$~\citep[see Fig.~8 of][]{Freedman2014}. This underestimation of the Rosseland mean opacity directly translates to an underestimation of the radiative gradient and to the appearance of a spurious deep radiative zone and a bias in our estimate of $P_{\rm R/C}$. In order to recover the behavior of the numerical model, we consider that a radiative zone that is in between two convectively unstable zone must have a radiative gradient smaller than $0.7$ times the adiabatic gradient, where the $0.7$ factor corresponds to the $\approx 30\%$ error in the fit of the Rosseland mean opacity. This reflects the sensitivity of the deep atmospheric structure to the Rosseland mean opacities calculations. As shown by~\citet{Freedman2014}, a variation of $\approx10-50\%$ is common between different calculations of the Rosseland mean opacities.   

Several trends concerning the depth of the radiative/convective boundary are apparent in Fig.~\ref{fig::PRC}. For a given effective temperature, the radiative/convective boundary is much deeper for irradiated planets than for non-irradiated planet ~\citep[as shown by \eg][]{Guillot2002}. Indeed the pressure of the radiative/convective boundary is almost constant while $\teff>\tint$ and decreases suddenly when $\teff\approx\tint$ and thus $\tmu\lesssim\tint$. The value of $P_{\rm R/C}$ also increases with gravity, which is directly linked to the decrease in the radiative gradient with gravity. Finally, we see that our analytical radiative/convective model properly predicts the depth of the radiative/convective boundary and its variation with gravity, internal temperature and effective temperature.

\begin{figure}[ht!]
\center
\includegraphics[width=\linewidth]{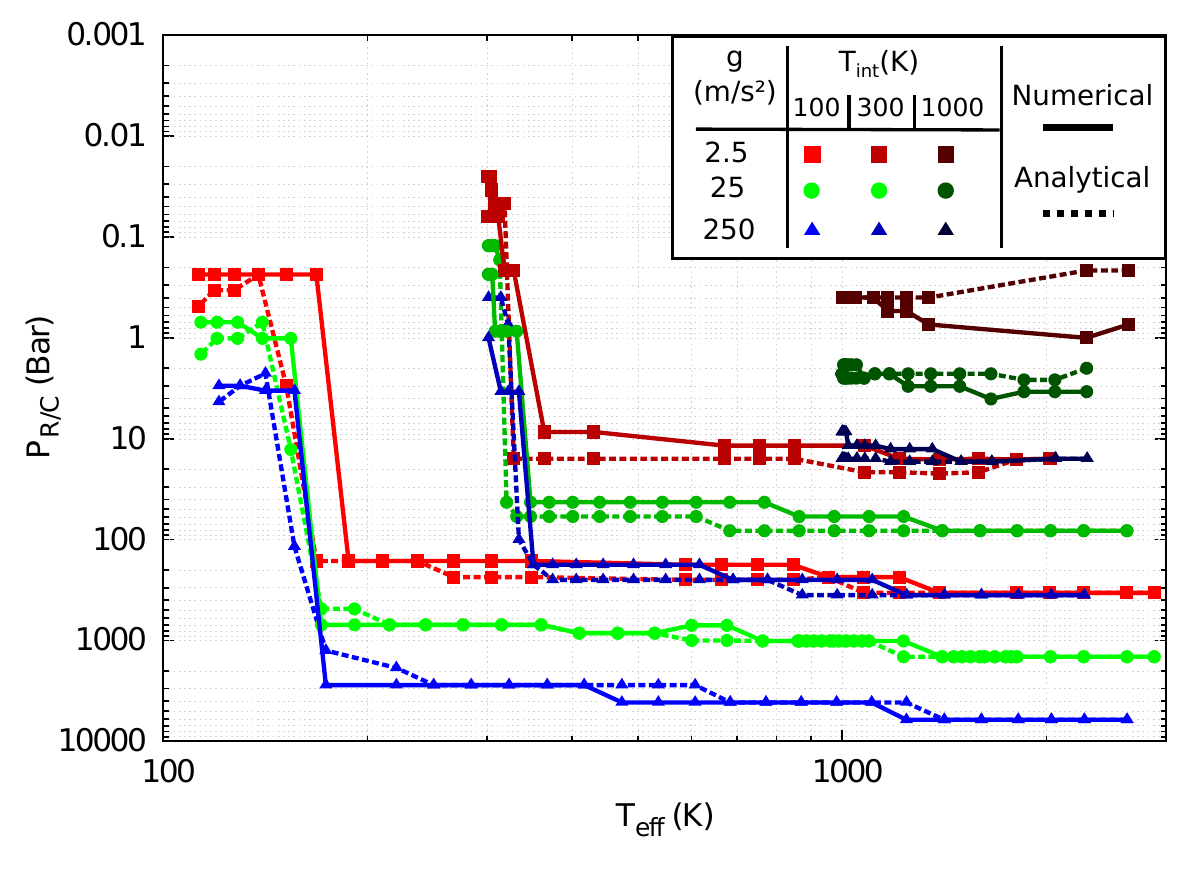}
\caption{{\tt Pressure of the radiative/convective boundary as a function of the effective temperature of the planet for different gravities and internal flux obtained with the numerical model (dashed lines) and with the analytical one (plain lines).}
}
\label{fig::PRC} 
\end{figure}

\subsection{Low irradiation planets and brown dwarfs}
\label{sec::VariousTint}
Gravitational contraction and deuterium burning can be a significant source of internal luminosity in young giant planets and brown dwarfs, respectively. This luminosity can overtake the stellar irradiation as the dominant heating source in the atmosphere. We calculated temperature/pressure profiles for planets with an internal temperature of $300\kelvin$ and $1000\kelvin$ and derived the same analytical models as in the case with $\tint=100\kelvin$ presented in Sect.~\ref{sec::Application}. Figs.~\ref{fig::AllProfiles-300K} and~\ref{fig::AllProfiles-1000K} show that at pressures of $1-100\,\bbar$ model D of of Sect.~\ref{sec::Application} -- derived considering an internal temperature of $100\kelvin$ -- correctly matches the numerical temperature/pressure profile for higher internal temperatures and can therefore be used as a boundary condition for internal structure model. {\tt In the the $T_{\rm int}=1000\,\K$ low gravity case we observe a discrepancy of $\approx 20\%$ between the numerical and the analytical model. This discrepancy is due to a the ionization of hydrogen not taken into account in our fit of the convective gradient.} The upper atmosphere is well represented as long as $\tint\lesssim\tmu$ (\ie $\teff\approx\tmu$). When the internal luminosity becomes the dominant energy source of the atmosphere, our analytical model becomes systematically hotter than the numerical model at low pressures with a discrepancy up to $40\%$ between the two models (see also Fig.~\ref{fig::ResultsAll} in appendix).
    \begin{figure}
\includegraphics[width=\linewidth]{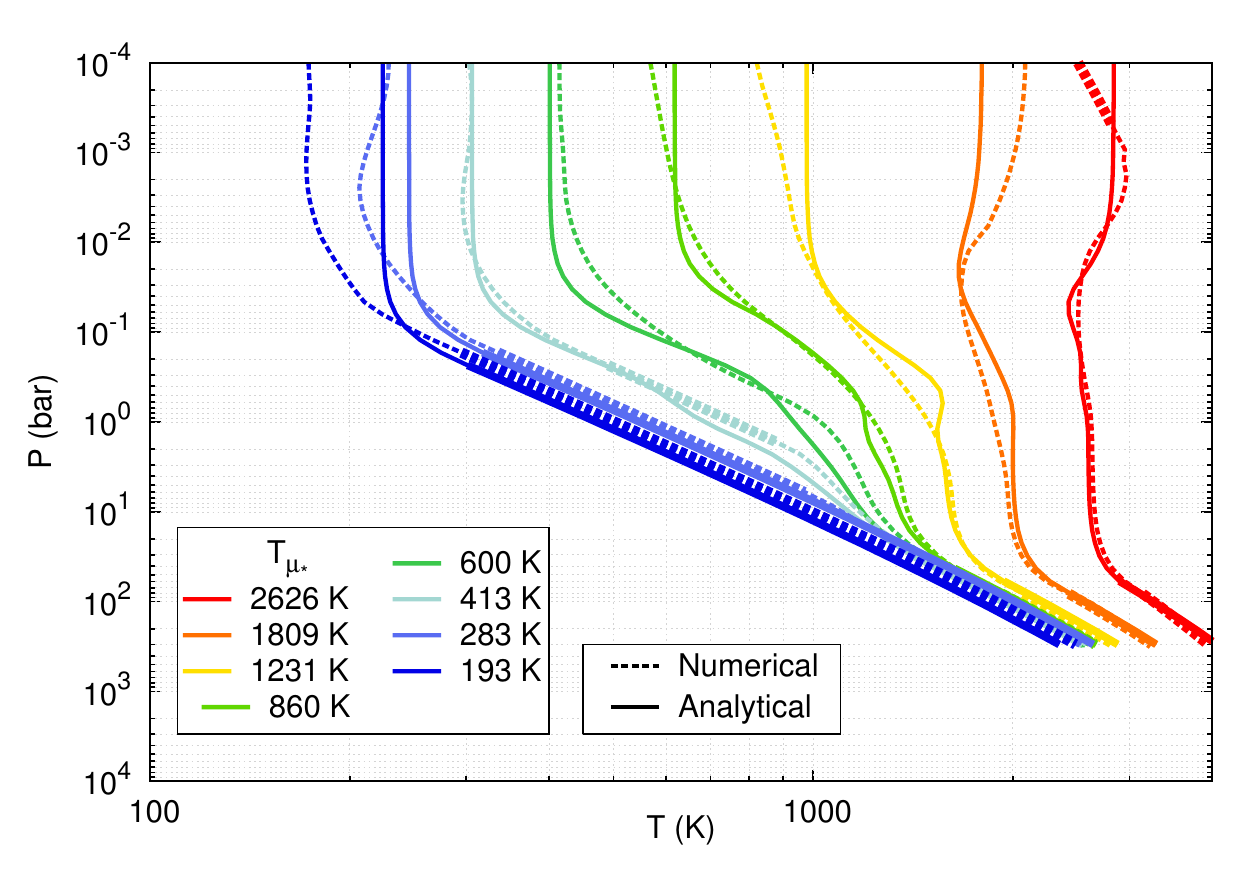}
\caption{Comparison of our numerical (dashed lines) and analytical (plain lines) solutions for $T_{\rm int}=300\,\kelvin$. For a full description see Fig.~\ref{fig::AllProfiles}.}
\label{fig::AllProfiles-300K}
\end{figure}

\begin{figure}
\includegraphics[width=\linewidth]{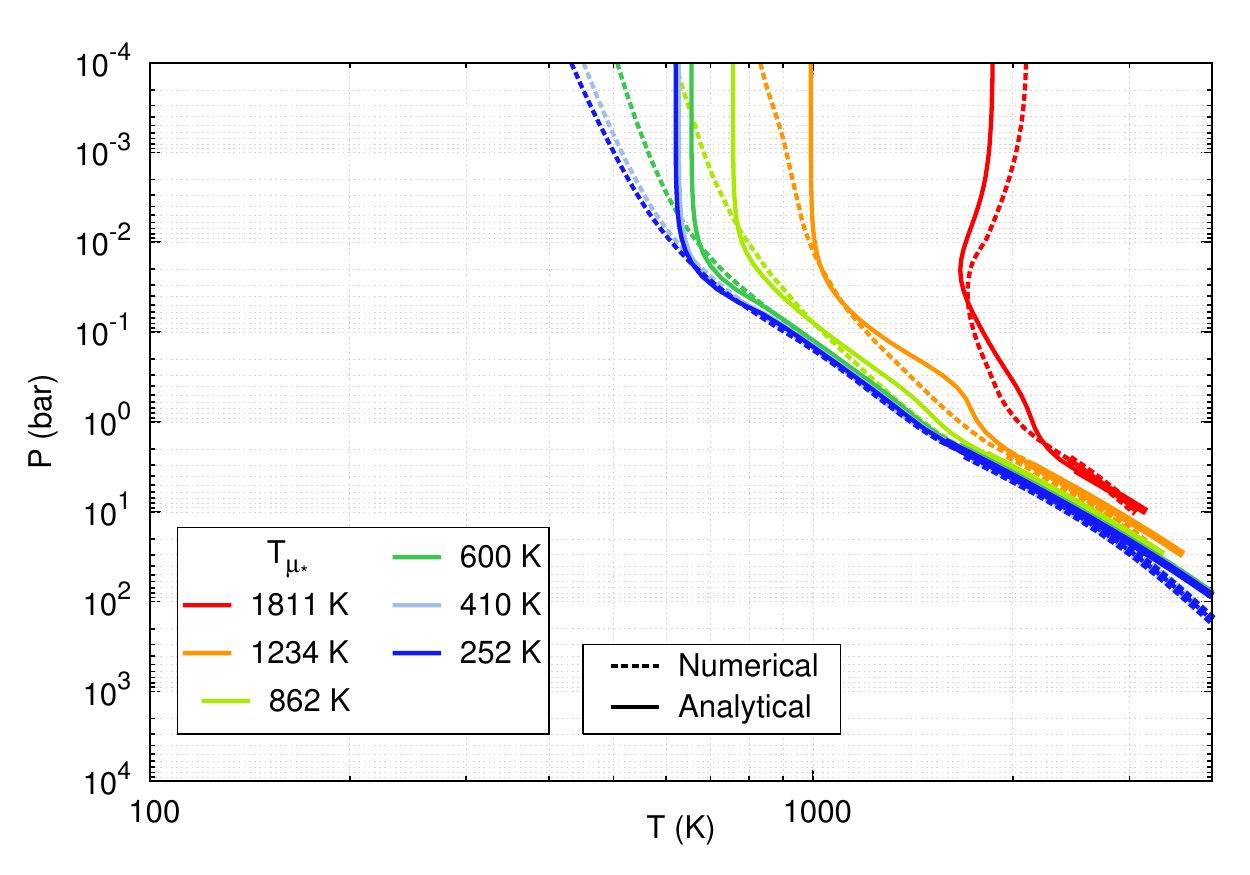}
\caption{Comparison of our numerical (dashed lines) and analytical (plain lines) solutions for $T_{\rm int}=1000\,\kelvin$. For a full description see Fig.~\ref{fig::AllProfiles}.}
\label{fig::AllProfiles-1000K}
\end{figure}

\subsection{Application to the atmospheres of solar-system giant planets}
In Fig.~\ref{fig::SS} we use model D of Sect.~\ref{sec::Application} without further adjustments to calculate the temperature-pressure profiles of the four giant planets of our solar-system. Our model pertains to cloudless, solar composition atmospheres, whereas Jupiter, Saturn, Uranus and Neptune have clouds and are all found to be significantly enriched in heavy elements. In spite of this, the analytical temperature profiles match the observations well (within 5 to 10\%) for pressures higher than about 0.05\,bar for Jupiter and Saturn and pressures higher than 0.1\,bar for Uranus and Neptune. Specifically, the temperature at 1 bar is found to be of 163 K, 131 K, 85 K and 77 K for the four planets. The corresponding temperatures inferred from observations are $165\pm 5$\,K, $135\pm 5$\,K, $76\pm 2$\,K, and $72\pm 2$\,K,respectively ~\citep[\eg][]{Lindal1992,Guillot2005}. Our model can therefore be used as a boundary condition for internal structure models of the solar-system giant planets. We predict that the radiative zone of Uranus extends down to $P\approx100\,\bbar$, what might affect its cooling history and outward thermal flux. Since our model cannot take into account the presence of a detached convective zone, it predicts an unphysical, higher-than-adiabatic gradient between 0.5 bar and 7 bar. For an internal flux equal to its observed $1-\sigma$ upper value, the convective zone extends continuously from the deep layers to the $\approx0.5\,\bbar$ level, leading to a more realistic temperature profile in Uranus' deep interior. A similar behavior is recovered when the opacities are increased by 50\%.

At low pressures, our model is systematically cooler than the observations, revealing the limits of our approach. In those cold planets, the photon deposition layer spans several orders of magnitudes in pressure and is poorly modeled by considering only two visible opacities. Moreover, our coefficients are constant with gravity whereas it is apparent from Fig.~\ref{fig::Results} that the dependance with gravity is stronger at low effective temperature.

\begin{figure}
\includegraphics[width=\linewidth]{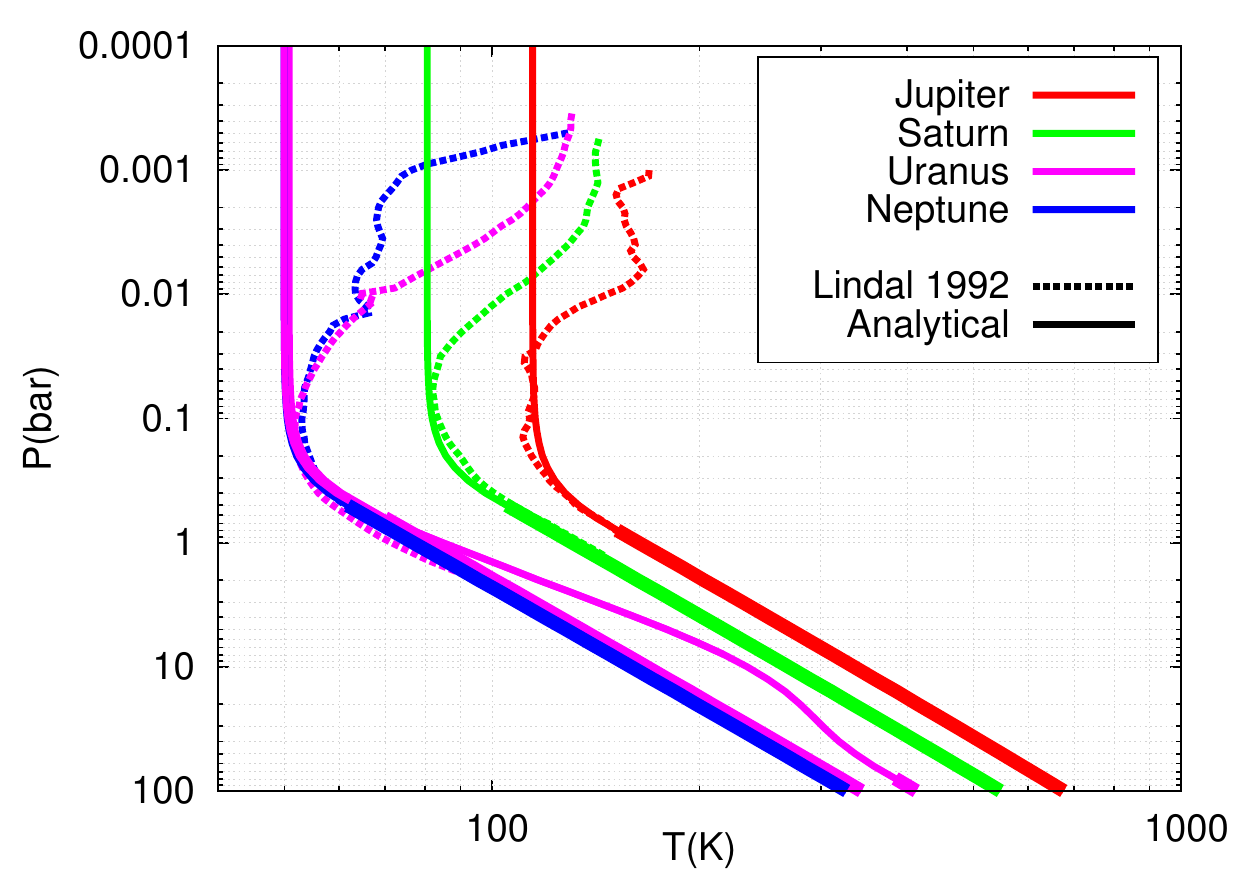}
\caption{{\tt Pressure temperature profiles of the four giant planets of the solar-system as derived from observations (dotted line, \citet{Lindal1992}) and from model D of section~\ref{sec::Application} (plain lines). The thin lines are radiative regions whereas the thick lines are convective regions. For Uranus we plot two profiles: one with the measured value of the internal temperature $T_{\rm int}=29.4\,K$ (the hotter profile) and one with the 1-$\sigma$ upper limit $T_{\rm int}=35.5\,K$ (the cooler profile). The fiducial model of Uranus have a deep radiative zone, it illustrates the difficulties of the method when a detached convective zone appears (here between 0.5 bar and 7 bar), since it is not taken into account by the analytical solution.}}
\label{fig::SS}
\end{figure}

\subsection{Recommended model}
\label{sec::resultssec}

When modeling gas giant planets of solar composition, we recommend the use of model D of Sect.~\ref{sec::Application}. This model uses the solution of the radiative transfer equation given by~\citetalias{Parmentier2014a} where the first five parameters describing the opacities are expressed as a function of the effective temperature (see Table~\ref{table::Models}) whereas the analytical Rosseland mean opacities are given by~\citet{Valencia2013}. 
Model D is fully analytical, yet achieves an overall accuracy of $10\%$ in temperature (at a given pressure) for irradiated giant planet atmospheres of solar composition with gravities in the range $2.5-250\,\meter\per\second^{2}$, internal temperatures of $100$ to $1000\,{\rm K}$ and effective temperatures from $100$ to $3000\,\kelvin$. {\tt When the internal flux dominates over the external flux (\ie\,$\tint>\tmu$), model D becomes less accurate in the medium and upper atmosphere with an error that can reach $\approx20\%$ and $\approx40\%$, respectively. In the deep atmospheres ($P\approx1--100\,\bbar$) it remains accurate and can be used as boundary condition for internal structure models.}

This accuracy is to be compared to that of simpler models. For example, models where the temperature is set to the effective temperature at $\tau=2/3$ and the profile is assumed to follow the diffusion approximation below (\ie~the so-called Eddington boundary condition). We calculated that this commonly used prescription~\citep[\eg][ among many others ]{Bodenheimer2003,Batygin2011} lead to an error in the temperature profile below the $\tau=2/3$ level on the order of $\approx30\%$ except fortuitously for $800K<\teff<1200K$ where the error is lower than $10\%$. Such an error on the boundary condition of interior models can strongly affects internal structure and planetary evolution calculations. Semi-grey model~\citep[\eg][]{Hansen2008,Guillot2010} cannot reach an accuracy better than $20\%$, even with adjusted variable opacity coefficients. 

A FORTRAN implementation of model D is available for download on the internet\footnote{http://www.oca.eu/parmentier/nongrey}.
{\tt
\section{The role of TiO and VO}
\label{sec::TiO}

The most irradiated planets have dayside atmospheric temperatures high enough such that, for a solar composition atmosphere some metal oxides such as titanium and vanadium oxides (TiO and VO, respectively), are chemically stable in gas phase~\citep{Lodders2002}. To this date, there is no firm direct detection of TiO at the atmospheric limb of an exoplanet~\citep[see][]{Desert2008,Huitson2013, Sing2013}. Condensation in a vertical cold trap~\citep{Spiegel2009}, in an horizontal cold trap~\citep{Parmentier2013}, dissociation by stellar radiation~\citep{Knutson2010} or the presence of a high C/O ratio~\citep{Madhusudhan2012a} have been proposed as mechanisms to deplete TiO and VO from the upper atmosphere of irradiated planets~\citep[see also the review in][]{Parmentier2014}.

Several studies~\citep[\eg][]{Hubeny2003,Fortney2008} show that, if present in solar abundance in the upper atmosphere of irradiated planets, titanium and vanadium oxides should change significantly the temperature structure of those atmospheres, creating a strong thermal inversion (also called stratosphere) at low pressures and reducing the temperature in the deep atmosphere. The signature of such a stratosphere have been searched for in the secondary eclipse spectra of a dozen of planets.  Although no unambiguous evidence for a stratosphere has yet been found, their presence have not been ruled out either~\citep{Hansen2014}. Here, we examine how TiO and VO opacities shape the atmospheric temperature profile of irradiated planets

\begin{figure}
  \includegraphics[width=\linewidth]{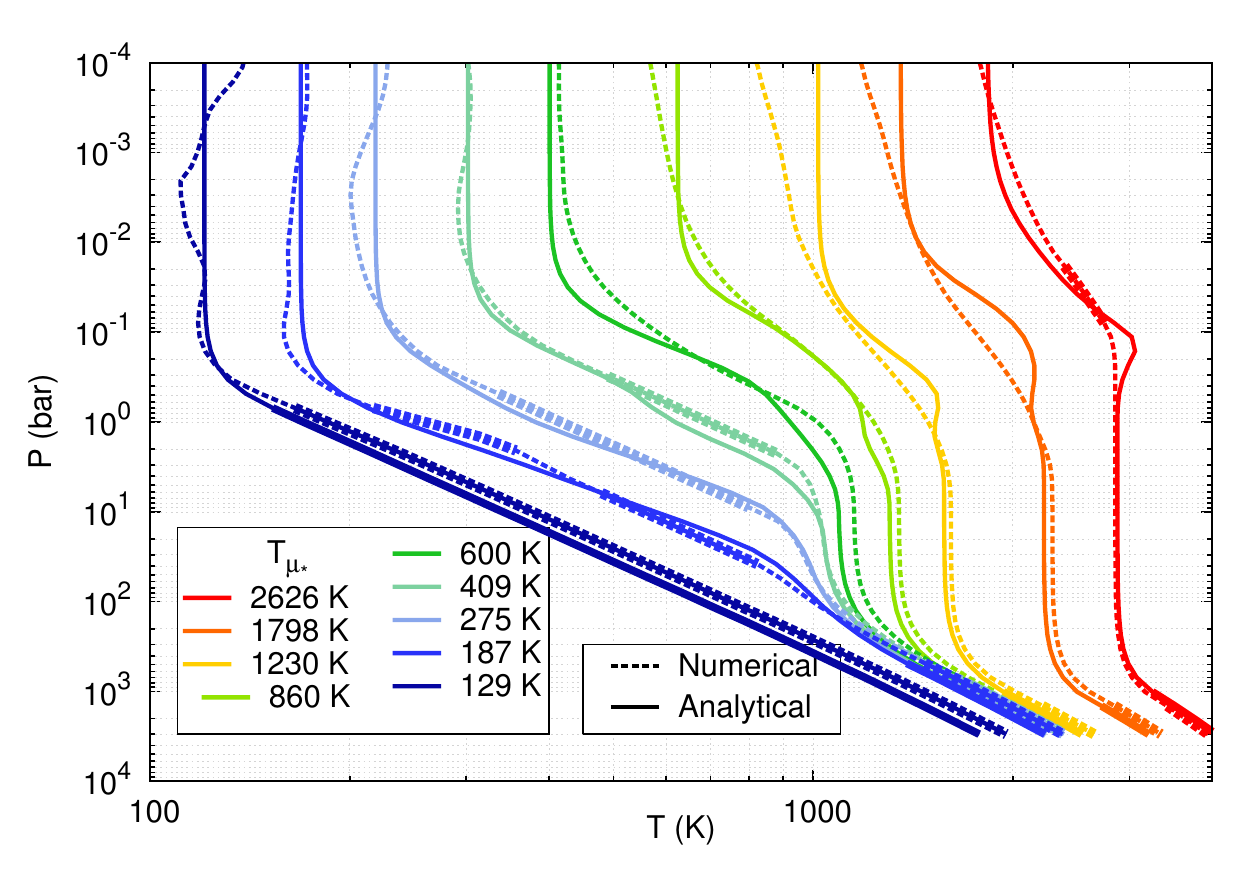}
\caption{Comparison of our numerical (dashed lines) and analytical (plain lines) solutions for an atmosphere without TiO/VO. For a full description see Fig.~\ref{fig::AllProfiles}.}
\label{fig::AllProfiles-NoTiO}
\end{figure}

\subsection{Retrieved coefficients}

\begin{figure*}
\includegraphics[width=\linewidth]{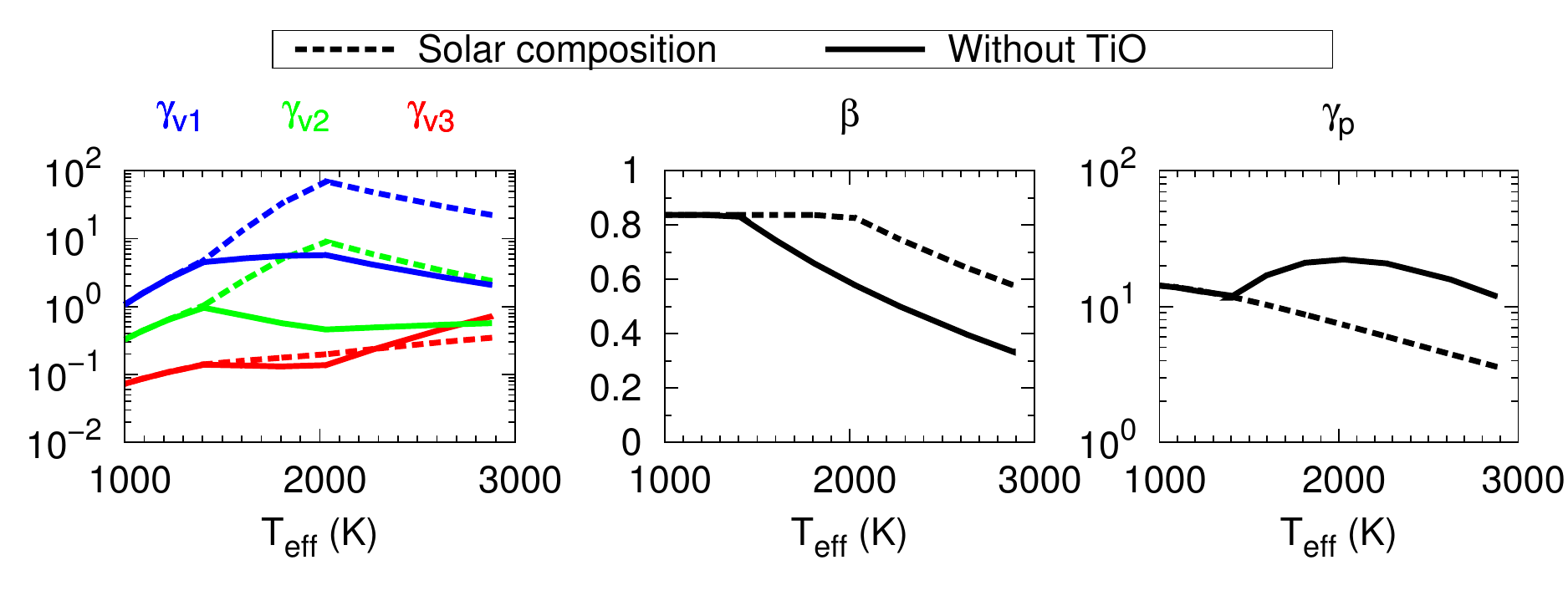}
\caption{{\tt Coefficients from our analytical model corresponding to the solar composition case (dashed lines, from Table~\ref{table::Models}) and the case where TiO and VO have been removed from the atmosphere (plain lines, Table~\ref{table::Models-NoTiO}).}}
\label{fig::CompTiO}
\end{figure*}

We calculated a grid of pressure/temperature profiles and derived the same analytical models as in the previous section in the case where TiO and VO have been removed from the whole atmosphere by any of the aforementioned processes. The resulting analytical model provides a good fit of the numerical profile, as seen in Fig~\ref{fig::AllProfiles-NoTiO} and quantified in Fig~\ref{fig::ResultsNoTiO} in appendix. The corresponding coefficients are presented in Table~\ref{table::Models-NoTiO} and compared with the solar composition case in Fig.~\ref{fig::CompTiO}.
 
As expected, the absence of TiO and VO changes significantly how the atmosphere absorbs the stellar irradiation: $\Gva$ and $\Gvb$ are 10 times smaller than in the solar composition case, showing that the first two thirds of the stellar energy are deposited at deeper levels than when TiO/VO are present. Conversely, $\Gvc$ remains almost constant : because the spectral width of the TiO band is smaller than the spectral width of the incoming stellar flux (see Fig~\ref{fig::Opacities-TiO} in appendix), part of the stellar flux remains unaffected by the presence of TiO and is absorbed at the same depth in the two cases. 

Usually not considered as important, the thermal coefficients are also affected by the presence of TiO and VO. $\beta$ is $\approx20\%$ smaller and $\Gp$ is $\approx10$ times higher than in the solar-composition case. Although TiO affects principally the optical region of the spectrum, hot Jupiters are hot enough that the local Planck function of their atmospheres can overlap with the TiO band. The broad-band opacities of TiO have two main effects: first, it fills the gap in the opacities seen between the Rayleigh scattering dominated opacities at small wavelength and the water bands beginning at $\approx1\,\mu\meter$, leading to a grayer atmosphere and thus a smaller value of $\gamma_{\rm p}$. Second, it changes the shape of the opacities: in the case without TiO, the opacities in the optical spectral range are dominated by  sodium and potassium lines, leading to smaller values of $\beta$ whereas in the solar composition case the optical opacities are dominated by the wide TiO band, leading to larger values of $\beta$.

\subsection{Effects of TiO and VO on the temperature profile}
We show that both the visible and the thermal coefficients are affected by the presence of TiO in the atmosphere. Thus both the greenhouse effect, the upper atmospheric cooling and the blanketing effect should contribute to the difference in the temperature profile between a solar-composition atmosphere and an atmosphere without TiO/VO. As shown by the blue and the red curves of Fig.~\ref{fig::NonGrey-TiO}, the disappearance of TiO and VO from the atmosphere reduces the temperature in the upper atmosphere (removing the temperature inversion) and increases the temperature in the deep atmosphere. 

We now wish to disentangle the contribution of the visible and the thermal effects in shaping the thermal profile between these two cases. We calculate an intermediate temperature profile in which the stellar flux is absorbed at the same Rosseland optical depth than in the solar-composition case but where TiO and VO have been removed from the thermal opacities. Thus, the grey model in Fig.~\ref{fig::NonGrey-TiO} have the same greenhouse effect than the solar-composition profile but have the same thermal effects than in the case without TiO/VO. This grey profile lay approximately at the midpoint between the profile with TiO/VO and the profile without TiO/VO. Non-grey thermal effects are therefore as important as non-grey absorption of the stellar light to set the difference between the cases with and without TiO/VO.

In the upper atmosphere, our intermediate profile does not show any temperature inversion ~\emph{although it absorbs the stellar irradiation with the same strength than in the solar-composition case}. The visible effects of TiO/VO alone cannot explain the presence of a temperature inversion in hot-Jupiters atmospheres. The inversion mainly happens because the broadband absorption of TiO increases the grayness of the opacities, reducing the ability of the upper atmosphere to cool down efficiently.

In the deep atmosphere, the absence of TiO and VO yields both a stronger greenhouse effect and a stronger blanketing effect. Both effects contribute equally to increase the deep temperature profile, leading to higher interior temperatures for the same global effective temperature and therefore to a slower evolution (\mbox{\citet{Parmentier2011a}}, \mbox{\citet{Budaj2012a}, \citet{Spiegel2013}}. In the absence of TiO and VO, hot Jupiters should be on average larger than if these gases are present.

\begin{figure}
  \includegraphics[width=\linewidth]{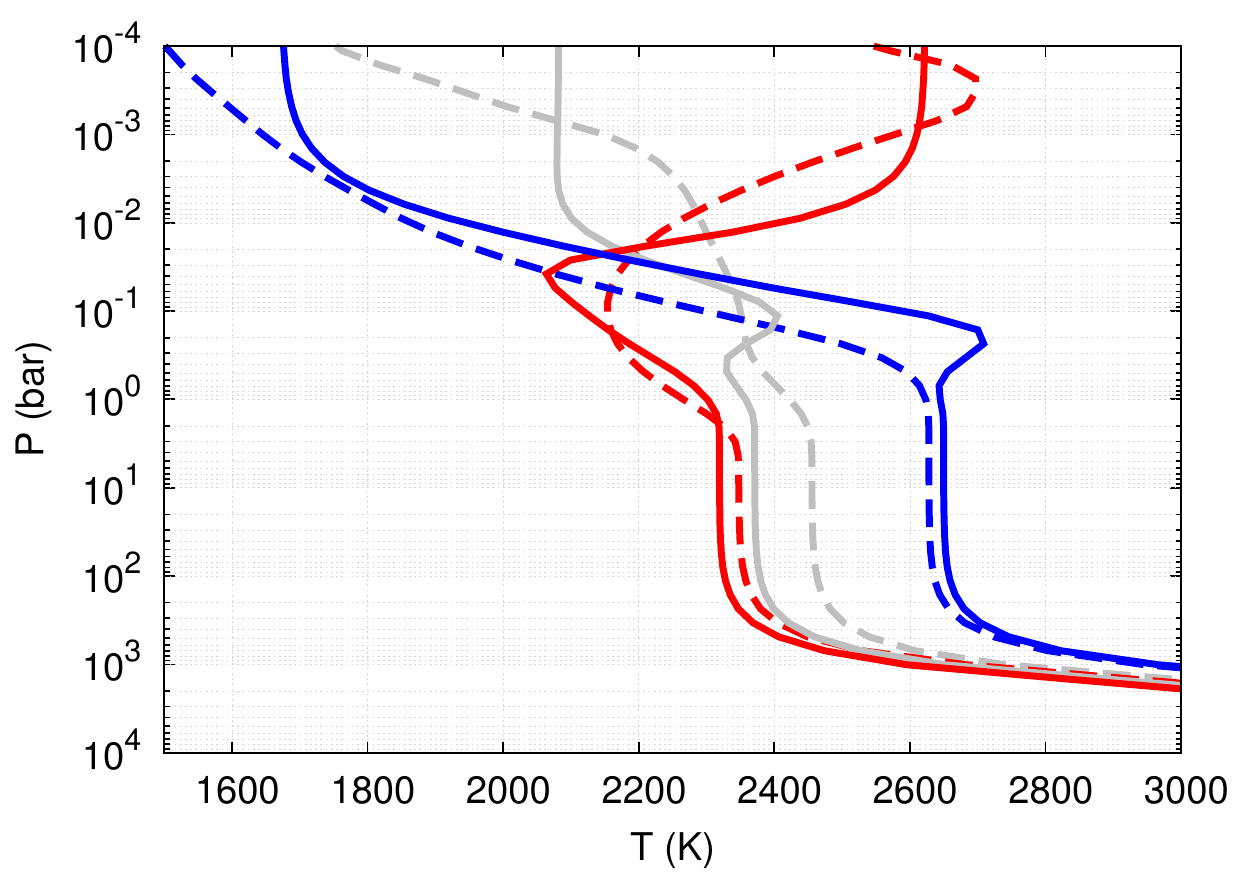}
\caption{{\tt Temperature-pressure profile for $T_{\rm eff}=2268\,\K$, $g=25\,\meter\per\second^{2}$ and $T_{\rm int}=100\,\K$ for a solar-composition atmosphere (red), an atmosphere where TiO and VO have been removed from the atmosphere (blue) and and atmosphere where the stellar irradiation is absorbed at the same optical depth as in the case with TiO/VO but where TiO/VO have been removed from the thermal opacities (grey). Our analytical model is shown as plain lines whereas the numerical solution is shown as dashed lines.}}
\label{fig::NonGrey-TiO}
\end{figure}
}

\section{Conclusion}

Analytical solutions of the radiative transfer equation, although derived using very restrictive (but necessary) approximations, offer a deep insight in the physical processes shaping the temperature profile of planetary atmospheres and can provide fast and roughly accurate solutions to be incorporated in more complex planetary models.

In this study we used hand-in-hand the analytical model derived in~\citetalias{Parmentier2014a}, that includes non-grey effects due to the visible and thermal opacities, and a state-of-the-art numerical model that solves the radiative transfer equation considering their full frequency and angular dependency. 

We first quantified the validity of the Eddington approximation. We showed that this approximation leads to errors in the temperature profile of at most $2\%$ in the grey case and $4\%$ in the non-grey case.

Planets with a thick atmosphere usually become convective below a certain depth. Thus, a common way to produce a radiative/convective temperature profile is to switch from a radiative solution to a convective solution whenever the Schwarzchild criterion is met, \emph{considering that the radiative solution remains unaffected by the presence of a convective zone below it.}
We showed that this approach is always valid in the grey case -- the error due to the Eddington approximation being frozen at the radiative/convective boundary and propagated along the convective zone. However, for non-grey atmospheric opacities, we showed that this method is valid only as long as the radiative/convective boundary remains in the optically thick layer of the atmosphere. When the radiative/convective boundary is in the optically thin region of the atmosphere, the radiative solution is very sensitive to the precise location of the radiative/convective boundary and this common approach can lead to relative errors of tens of percentage points when estimating the upper, radiative, atmospheric temperatures.

We showed that semi-grey effects (greenhouse and anti-greenhouse effects) are not sufficient to explain the atmospheric temperature profiles calculated with the full frequency dependent opacities and that non-grey thermal effects need to be taken into account. We provided a reliable method to obtain the visible coefficients of our analytical model directly from the opacities and explored how the thermal coefficients could also be directly derived from the knowledge of the line-by-line atmospheric opacities.

In particular, we showed that the presence of TiO can warm up the upper atmosphere and cool down the deep atmosphere not only because it absorbs a significant amount of stellar irradiation in the upper atmosphere, but also because its broad band opacity reduces the non-grey thermal blanketing effect. The presence of a thermal inversion in hot-Jupiters induced by an extra absorber in the optical opacities cannot be explained solely by an increased absorption of the stellar light and depends mostly on the ability of the atmosphere to cool down via the non-grey thermal effects. 

Finally, using an \emph{a-priori} determination of the visible coefficients and an \emph{a-posteriori} determination of the thermal coefficients, we provide a fully analytical model for solar composition optically thick atmospheres. This model agrees with the numerical calculations within $10\%$ over a wide range of gravities and effective temperatures. Our model leads to a much better estimate of the deep temperature profile than the previous analytical estimates. Therefore, when modeling the atmospheric structure of giant planets, we recommend the use of Model D described in Sect.~\ref{sec::resultssec} that uses the analytical expressions derived in~\citetalias{Parmentier2014a} with the first five parameters given in Table~\ref{table::Models} for a solar-composition atmosphere and by Table~\ref{table::Models-NoTiO} in the case where TiO and VO have been removed from the atmosphere. The Rosseland mean opacities of model D are given by the functional form of~\citet{Valencia2013} and the convective gradient follows Eq.~\eqref{eq::ConvGrad}. For convenience, we provide an implementation in FORTRAN of our model in the CDS and at the address \emph{http://www.oca.eu/parmentier/nongrey}.

\section*{Acknowledgments}
We acknowledge the anonymous referee for his general comments that increased the quality of this manuscript. This work was performed in part thanks to a joint Fulbright Fellowship to V.P. and T.G. The whole project would not have been possible without the help and support of Douglas Lin. We also acknowledge University of California Santa Cruz for hosting us while this work was carried out. During the last part of this work, V.P. was under contract with the Jet Propulsion Laboratory (JPL) funded by NASA through the Sagan Fellowship Program executed by the NASA Exoplanet Science Institute.

\bibliography{../../../../../Biblio/biblio.bib}
\clearpage
\begin{appendix}
\section{Additional material}
\label{Sec::AppendixOpacities}
The opacities in the form of k-coefficients used in the numerical model and discussed in Sect.~\ref{sec::Application} are presented in Fig.~\ref{fig::Opacities-TiO}. 

The analytical model adjusted to match the temperature/pressure profile of an atmosphere without TiO, discussed in Sect.~\ref{sec::TiO} is presented in Fig.~\ref{fig::ResultsNoTiO} and Table~\ref{table::Models-NoTiO}.

The effect of a strong internal luminosity on the analytical model, discussed in Sect.~\ref{sec::VariousTint} is presented if Fig.~\ref{fig::ResultsAll}.

\begin{figure*}[!ht]
   \begin{minipage}[c]{.45\linewidth}
\includegraphics[width=\linewidth]{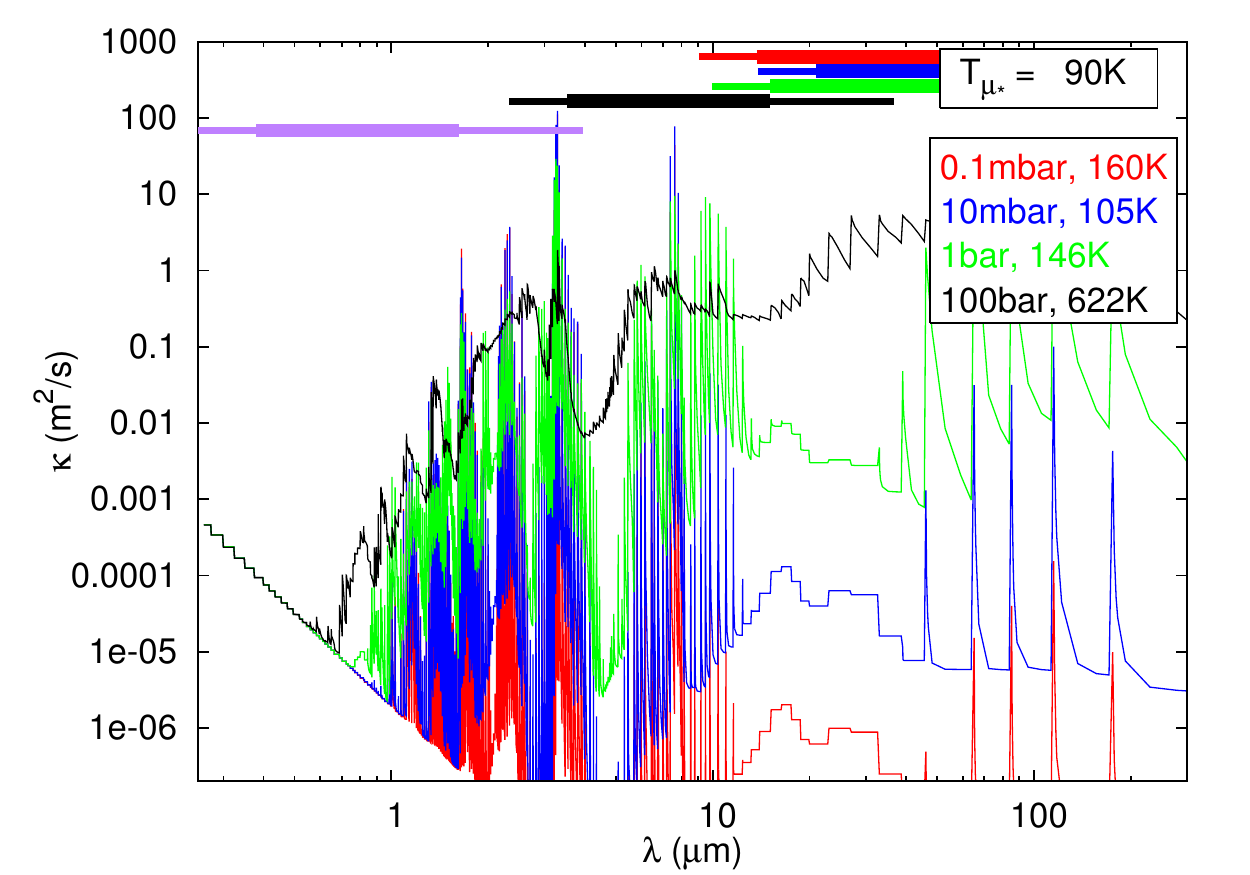}
\includegraphics[width=\linewidth]{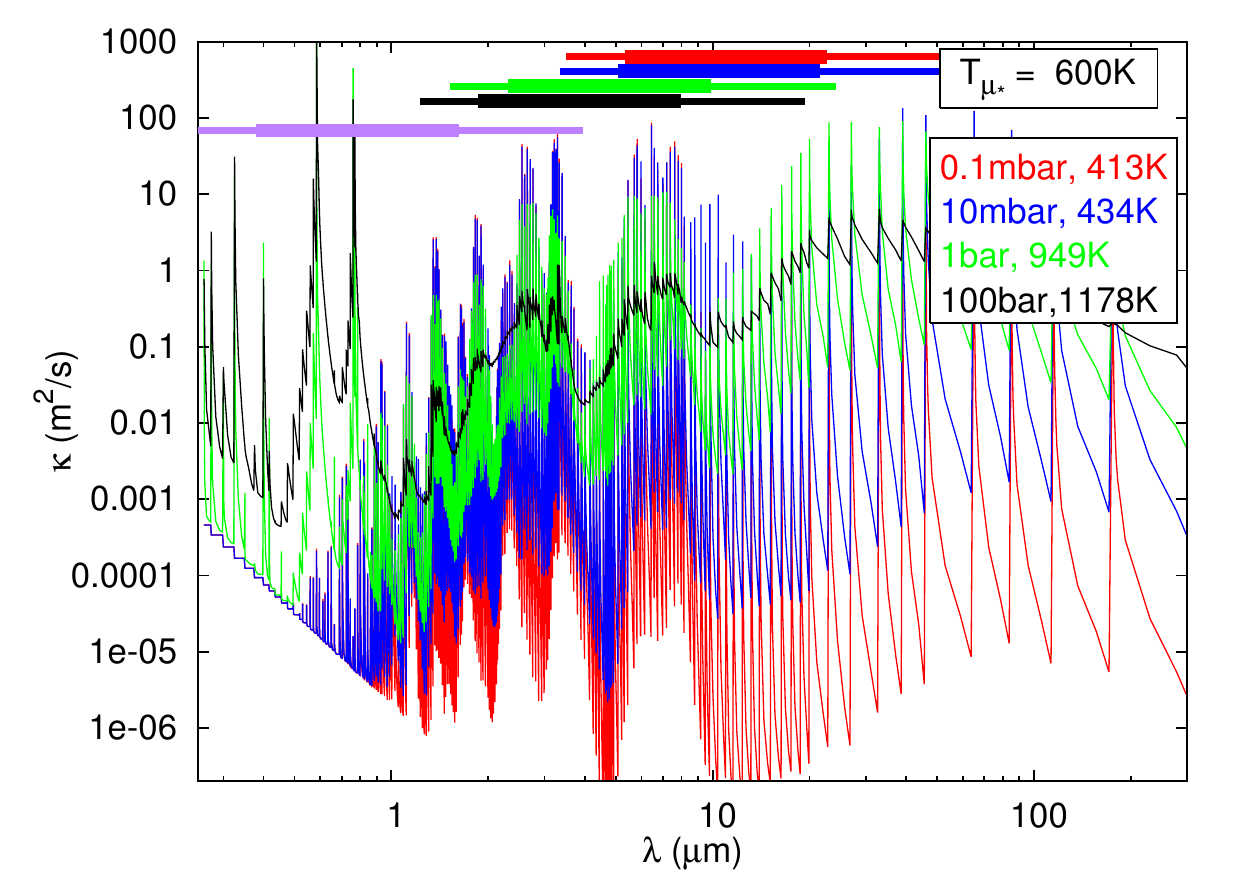}
\includegraphics[width=\linewidth]{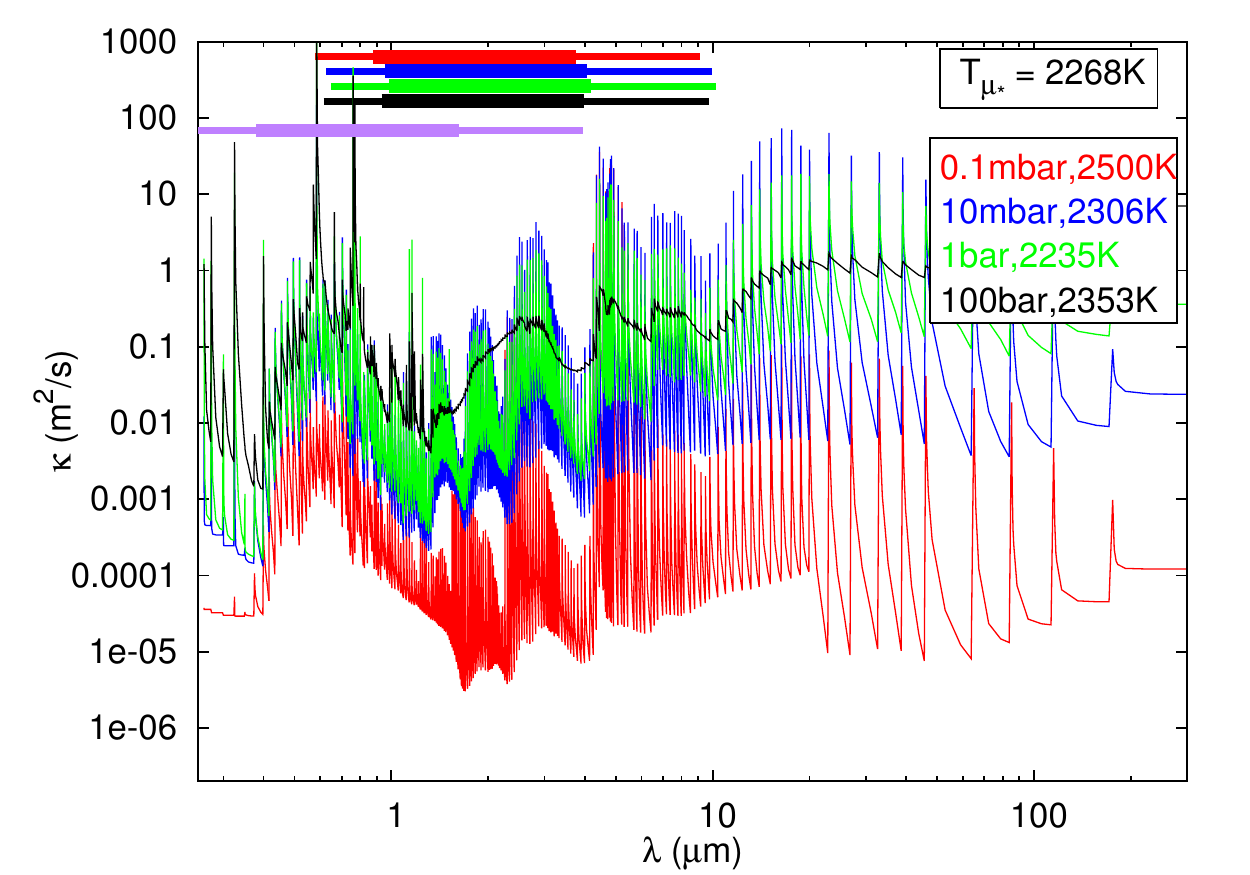}
   \end{minipage} \hfill
   \begin{minipage}[c]{.45\linewidth}
\includegraphics[width=\linewidth]{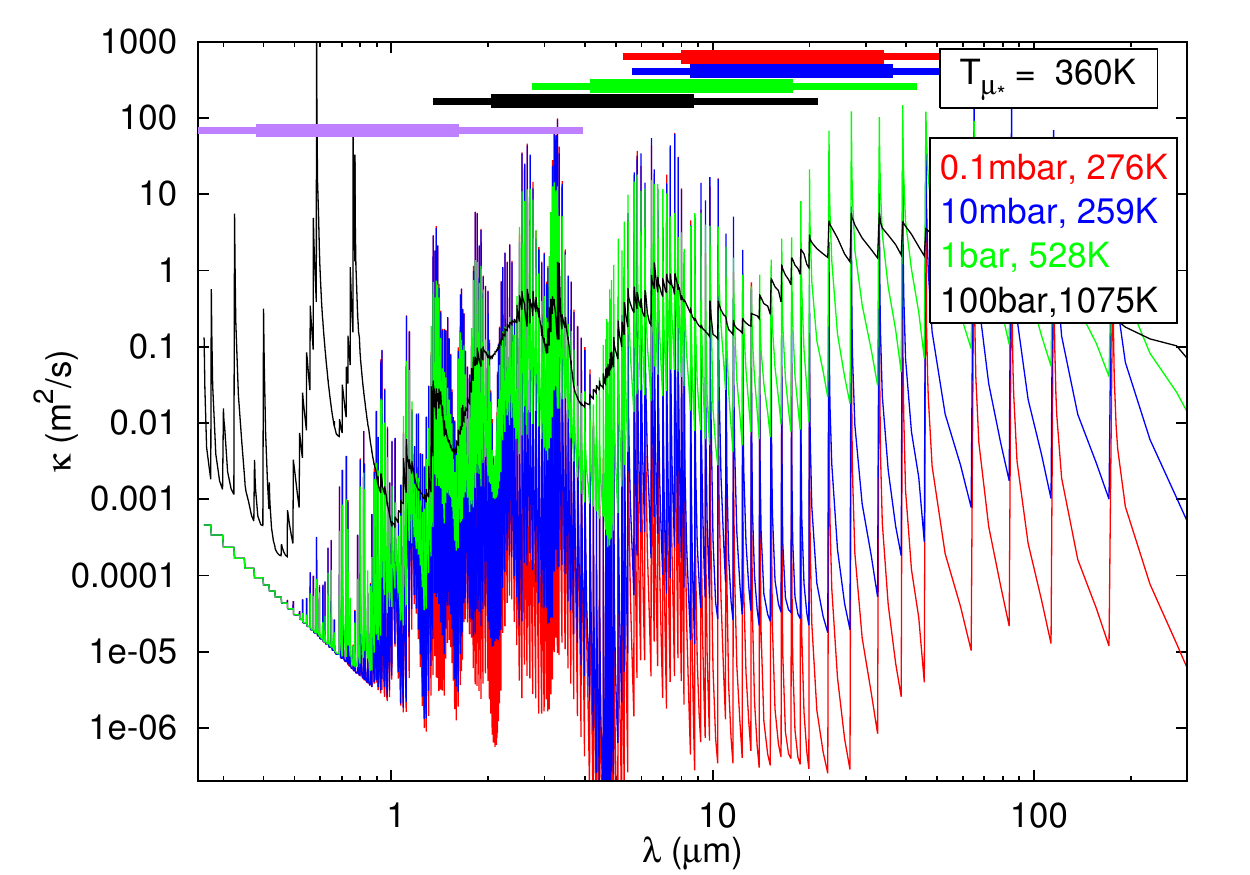}
\includegraphics[width=\linewidth]{./images/Opacities-TiO/Opacities-A-053_G25_T100_z0.pdf}
\includegraphics[width=\linewidth]{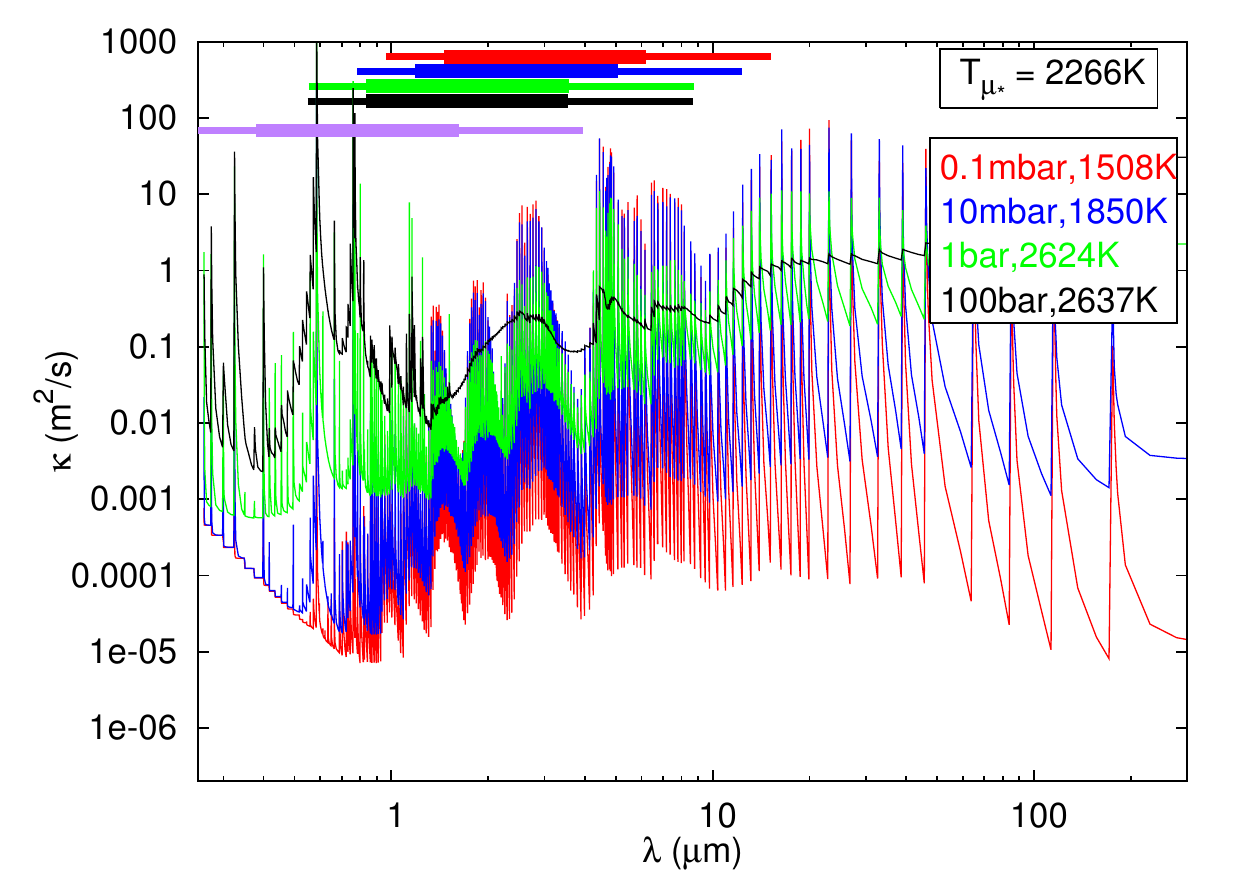}
   \end{minipage}
  \caption{Opacities from \cite{Freedman2008} organized as k-coefficient inside each bin of wavelength atmospheres with different irradiation. The first five panels are for solar-composition atmospheres whereas the bottom right panel is for an atmospheres depleted in TiO/VO. The different colors are for different temperature and pressure taken along the corresponding numerical P-T profile. The thick bars on top represents the wavelength range where $90\%$ of the thermal flux is emitted, the thin bars where $99\%$ of the thermal flux is emitted. We used $\tint=\unit{100}K$, $\mu_{*}=1/\sqrt{3}$ and $g=\unit{25}\meter\per\second\squared$.}
   \label{fig::Opacities-TiO}
\end{figure*}

\begin{figure*}[ht!]
\center
\includegraphics[width=0.82\linewidth]{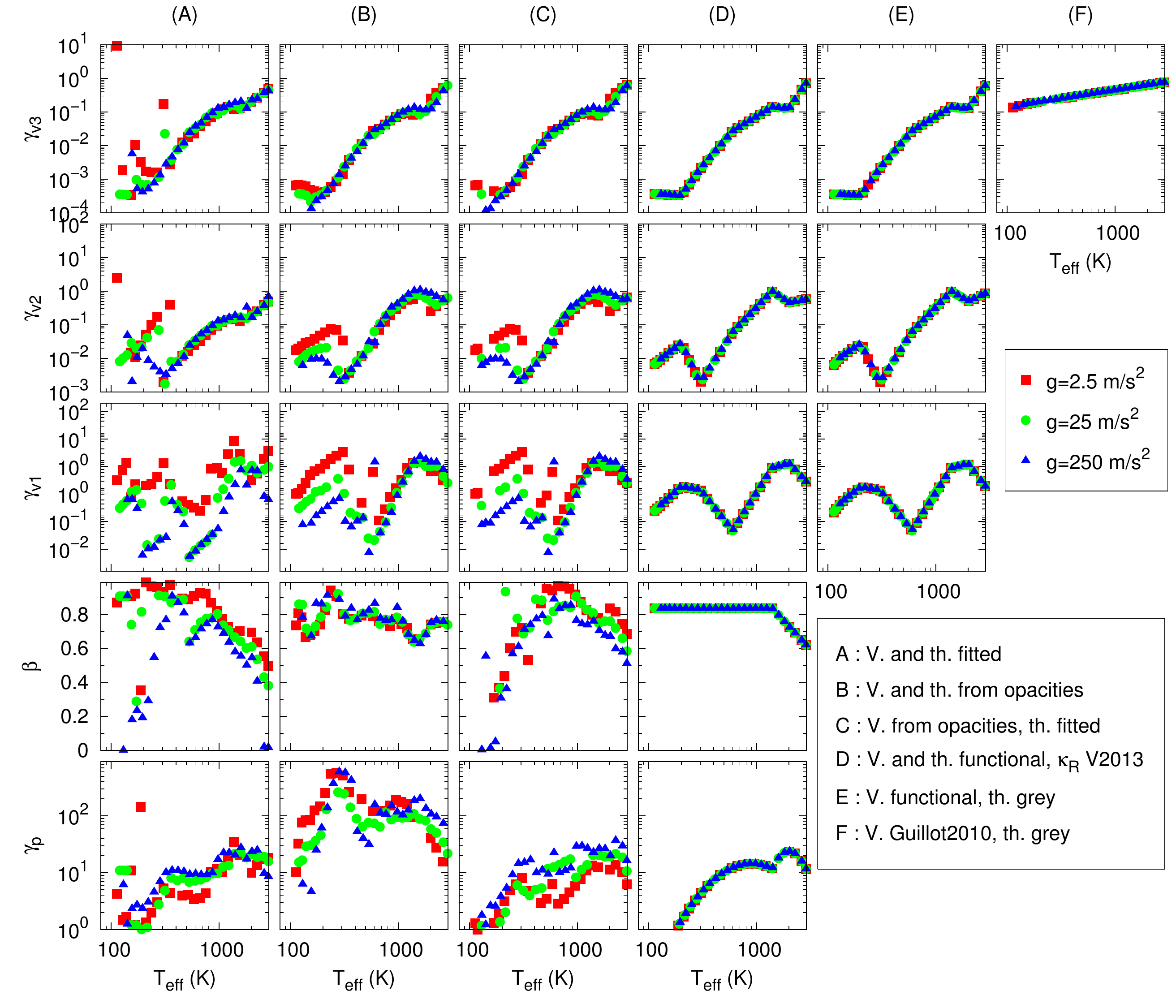}
\includegraphics[width=0.82\linewidth]{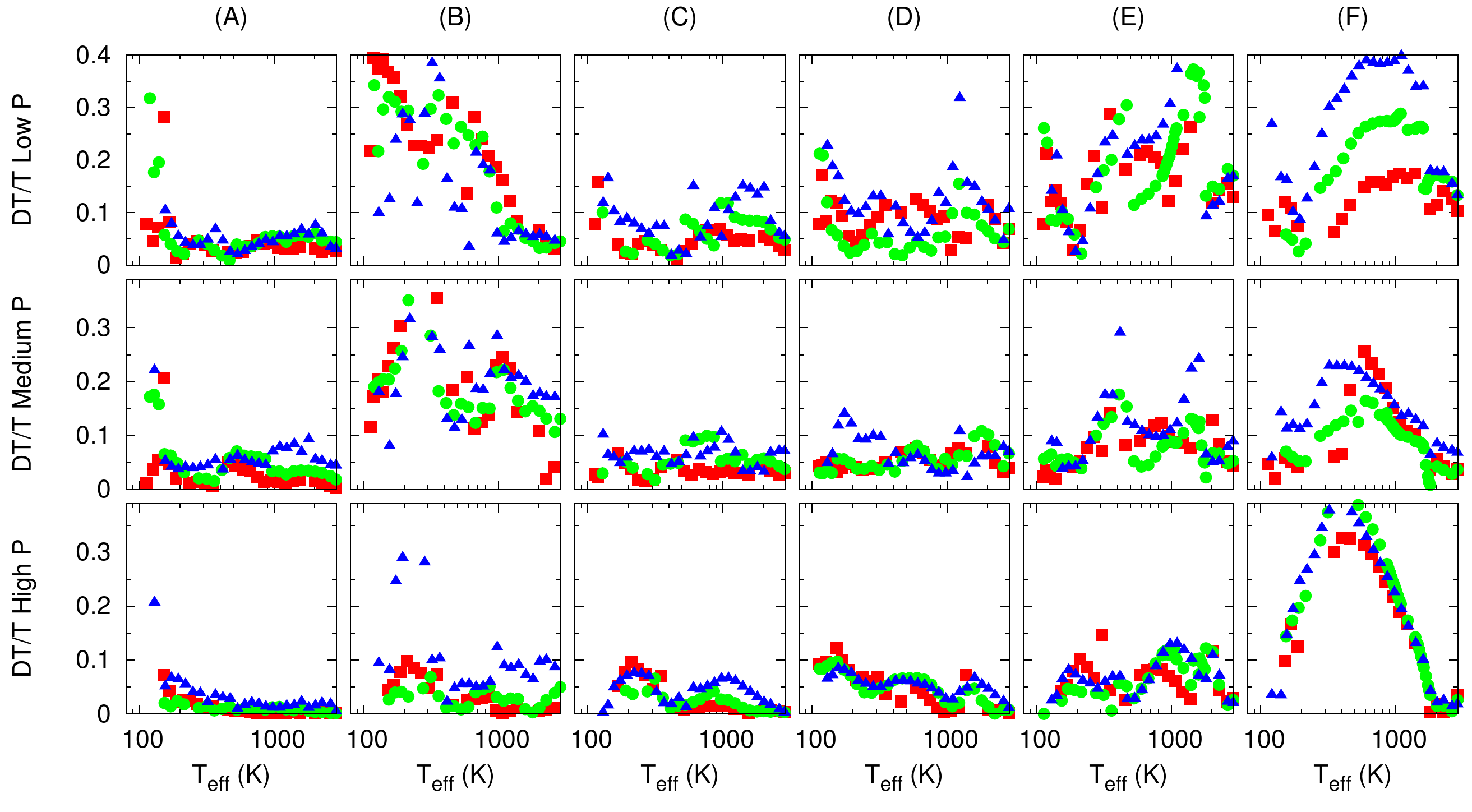}
\caption{
\emph{Top panel}: coefficients obtained for the six different models described in Sect.~\ref{sec::Application} as a function of the irradiation temperature for planets with different gravities and an internal temperature of $100\kelvin$. TiO and VO have been artificially removed from the atmosphere.
\emph{Bottom panel}: Mean relative difference between the numerical and the analytical model for the six different models described in Sect.~\ref{sec::Application}. The first line is the mean difference for $\unit{10^{-4}}\bbar<P<\unit{10^{-2}}\bbar$, the second line for $\unit{10^{-2}}\bbar<P<\unit{10^{0}}\bbar$ and the third line for $\unit{10^{0}}\bbar<P<\unit{10^{2}}\bbar$. In terms of Rosseland optical depth, the low pressure zone corresponds to the optically thin part of the atmosphere with $10^{-8}(25\,{\rm m s^{-2}}/g)\lesssim\tau_{\rm R}\lesssim10^{-2}(25\,{\rm m s^{-2}}/g)$. The medium pressure zone corresponds to the transition from optically thin to optically thick with $10^{-4}(25\,{\rm m s^{-2}}/g)\lesssim\tau_{\rm R}\lesssim10(25\,{\rm m s^{-2}}/g)$. The high pressure zone corresponds to the optically thick part of the atmosphere with $(25\,{\rm m s^{-2}}/g)\lesssim\tau_{\rm R}\lesssim10^{4}(25\,{\rm m s^{-2}}/g)$.}

\label{fig::ResultsNoTiO} 
\end{figure*}

\begin{center}
\begin{table*}

\caption{Functional form of the coefficients of the analytical model of~\citetalias{Parmentier2014a} valid for atmospheres where TiO and VO have been removed. We use $X=\log_{10}(\teff)$}
\centering
\ra{1.7} 
\begin{tabular}{>{\centering}m{1.5cm} >{\centering} m{1.9cm} >{\centering} m{1.9cm} >{\centering} m{1.9cm} >{\centering} m{1.9cm} >{\centering} m{1.9cm}  >{\centering} m{1.9cm}  >{\centering} m{1.9cm}}

\toprule 
Coefficient & Expression & $\teff<200\,\K$ & $200\,\K<\teff<300\,\K$ & $300\,\K<\teff<600\,\K$ & $600\,\K<\teff<1400\,\K$& $1400\,\K<\teff<2000\,\K$ & $\teff>2000\,\K$\\ 
\midrule
\multirow{2}{*}{$\log_{10}(\Gvc)$}&\multirow{2}{*}{$a+bX$}&$a=-3.03$&$a=-13.87$&$a=-11.95$&$a=-6.97$&$a=0.02$&$a=-16.54$\\
&&$b=-0.2$&$b=4.51$&$b=3.74$&$b=1.94$&$b=-0.28$&$b=4.74$\\\hline
\multirow{2}{*}{$\log_{10}(\Gvb)$}&\multirow{2}{*}{$a+bX$}&$a=-7.37$&$a=13.99$&$a=-15.18$&$a=-10.41$&$a=6.96$&$a=-2.4$\\
&&$b=2.53$&$b=-6.75$&$b=5.02$&$b=3.31$&$b=-2.21$&$b=0.62$\\\hline
\multirow{2}{*}{$\log_{10}(\Gva)$}&\multirow{2}{*}{$a+bX$}&$a=-5.51$&$a=1.23$&$a=8.65$&$a=-12.96$&$a=-1.68$&$a=10.37$\\
&&$b=2.48$&$b=-0.45$&$b=-3.45$&$b=4.33$&$b=0.75$&$b=-2.91$\\\hline
\multirow{2}{*}{$\beta$}&\multirow{2}{*}{$a+bX$}&$a=0.84$&$a=0.84$&$a=0.84$&$a=0.84$&$a=3$&$a=3$\\
&&$b=0$&$b=0$&$b=0$&$b=0$&$b=-0.69$&$b=-0.69$\\\hline
$\log_{10}(\Gp)$&$aX^{2}+bX+c$&\multicolumn{4}{c}{\hfill$a=-2.36$, $b=13.92$, $c=-19.38$\hfill $\vert$ }&\multicolumn{2}{c}{ $a=-12.45$, $b=82.25$, $c=-134.42$\hfill}\\\hline
\bottomrule 
\label{table::Models-NoTiO}

\end{tabular}
 
 \end{table*}
\end{center}

\begin{figure*}[ht!]
\center
\includegraphics[width=0.79\linewidth]{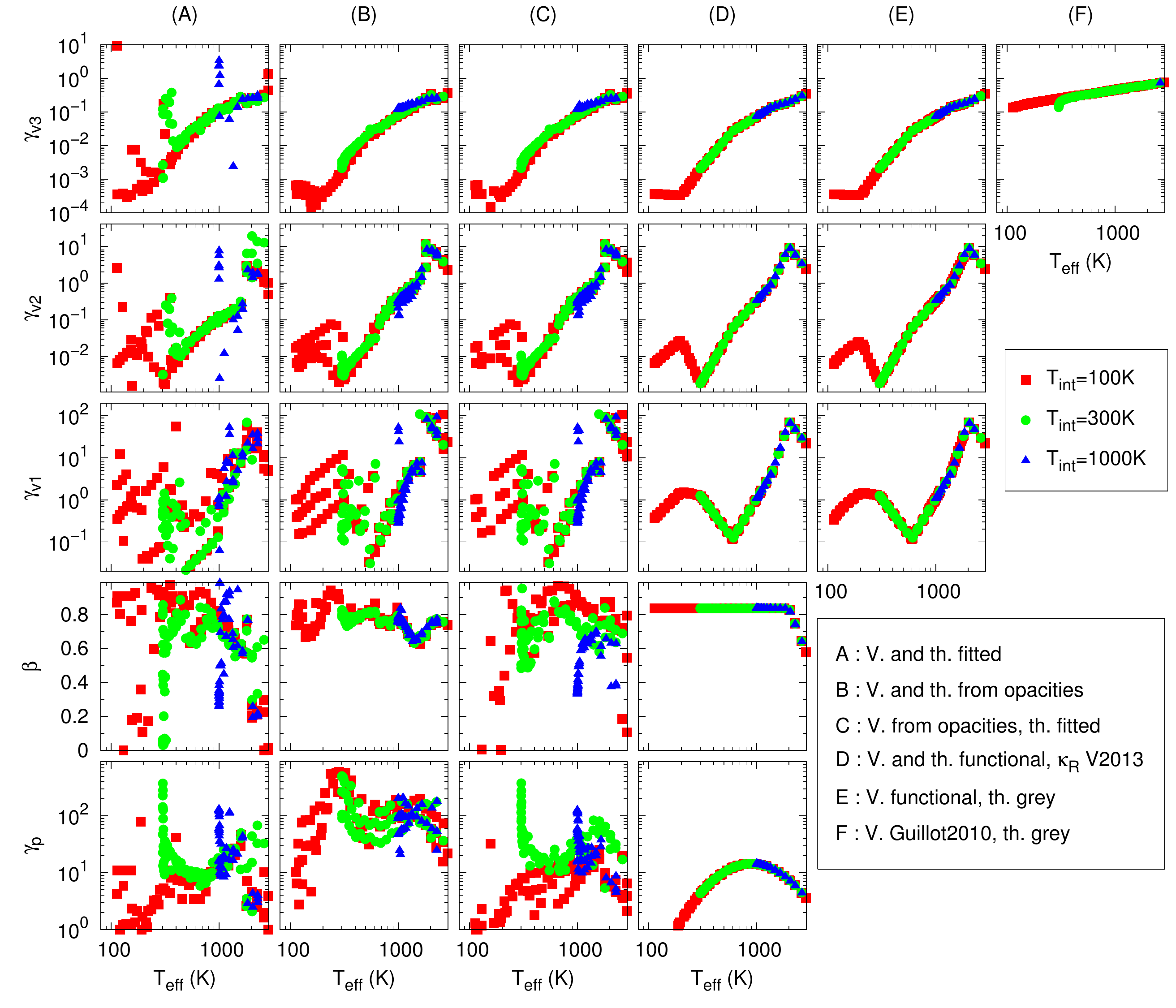}
\includegraphics[width=0.79\linewidth]{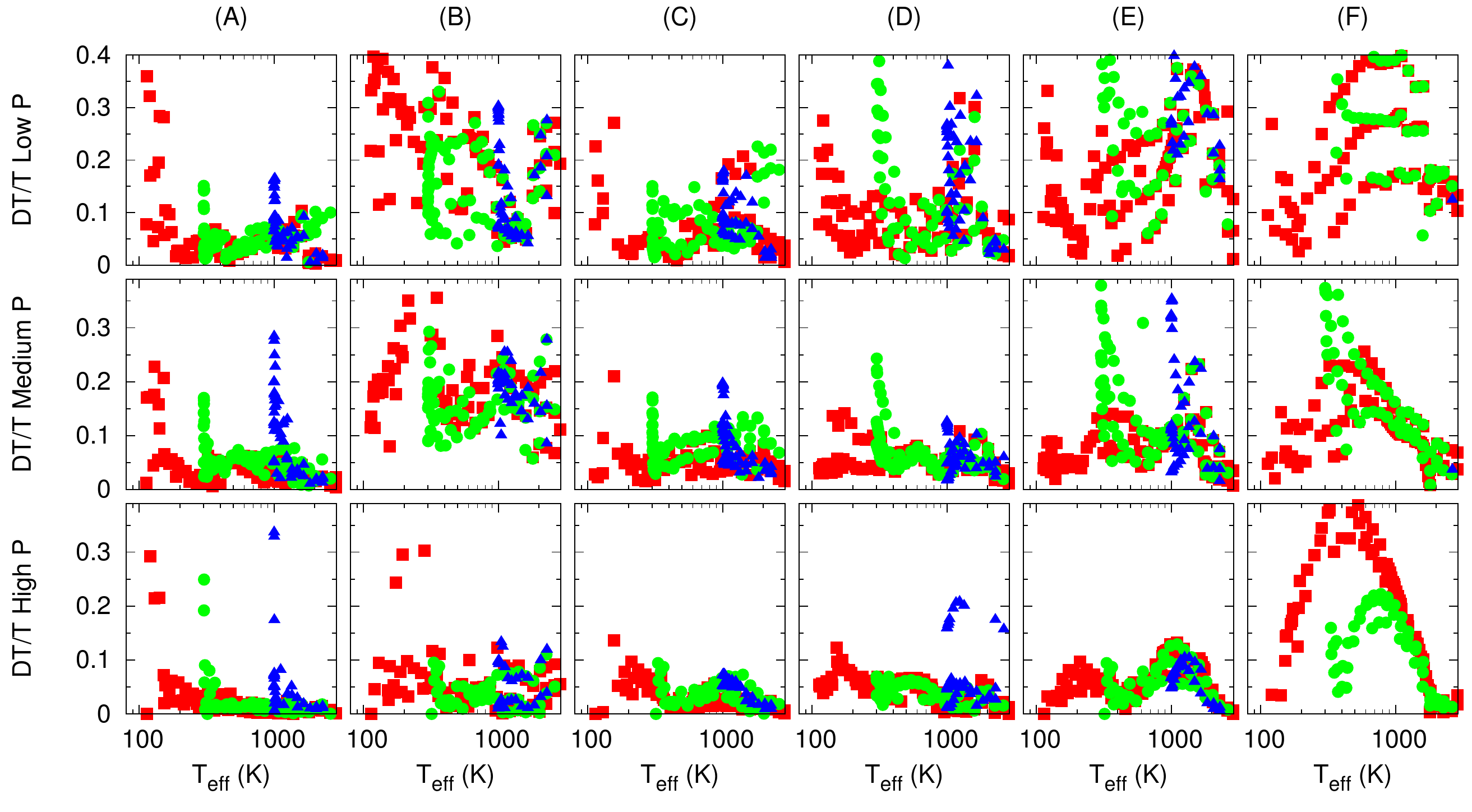}
\caption{
\emph{Top panel}: coefficients obtained for the six different models described in Sect.~\ref{sec::Application} as a function of the irradiation temperature for planets with a solar composition atmosphere with different gravities and an internal temperature of $100\kelvin$, $300\kelvin$ and $1000\kelvin$. The model of column D uses the functional form of the coefficients derived in the case $\tint=100\kelvin$ only.
\emph{Bottom panel}: Mean relative difference between the numerical and the analytical model for the six different models described in Sect.~\ref{sec::Application}. The first line is the mean difference for $\unit{10^{-4}}\bbar<P<\unit{10^{-2}}\bbar$, the second line for $\unit{10^{-2}}\bbar<P<\unit{10^{0}}\bbar$ and the third line for $\unit{10^{0}}\bbar<P<\unit{10^{2}}\bbar$. In terms of Rosseland optical depth, the low pressure zone corresponds to the optically thin part of the atmosphere with $10^{-8}(25\,{\rm m s^{-2}}/g)\lesssim\tau_{\rm R}\lesssim10^{-2}(25\,{\rm m s^{-2}}/g)$. The medium pressure zone corresponds to the transition from optically thin to optically thick with $10^{-4}(25\,{\rm m s^{-2}}/g)\lesssim\tau_{\rm R}\lesssim10(25\,{\rm m s^{-2}}/g)$. The high pressure zone corresponds to the optically thick part of the atmosphere with $(25\,{\rm m s^{-2}}/g)\lesssim\tau_{\rm R}\lesssim10^{4}(25\,{\rm m s^{-2}}/g)$ When $\teff\approx\tint$ the low pressures are not properly represented by our model (see Sect.~\ref{sec::VariousTint} for more details).}
\label{fig::ResultsAll} 
\end{figure*}

\end{appendix}

\end{document}